\documentclass[a4paper]{article}
\usepackage[utf8x]{inputenc}
\usepackage[left=3cm,right=3cm,top=3cm,bottom=3cm]{geometry}
\usepackage[linktoc=all,hidelinks]{hyperref}
\usepackage{cite}

\usepackage[T1]{fontenc}
\usepackage{amsmath}
\usepackage{amssymb}
\usepackage{esint}
\usepackage{amsfonts}
\usepackage{graphicx,color}
\usepackage{slashed}
\usepackage{young}
\usepackage[vcentermath]{youngtab}
\usepackage{tensor}
\usepackage{etoolbox}
\usepackage{bbm}
\usepackage{soul}
\usepackage{ulem}

\usepackage{epic,eepic,pstricks}
\usepackage{mathrsfs}

\usepackage{bbm}

\definecolor{darkgreen}{rgb}{0.2,0.7,0.2}

\newcommand{\ie}{i.e.~}

\newcommand{\be}{\begin{equation}}
\newcommand{\ee}{\end{equation}}
\newcommand{\bea}{\begin{eqnarray}}
\newcommand{\eea}{\end{eqnarray}}

\newcommand*\xbar[1]{%
  \hbox{%
    \vbox{%
      \hrule height 0.5pt 
      \kern0.3ex
      \hbox{%
        \kern-0.0em
        \ensuremath{#1}%
        \kern-0.0em
      }%
    }%
  }%
}

\numberwithin{equation}{section}

\begin{document}

\thispagestyle{empty}
\begin{center}

\vspace*{50pt}
{\LARGE \bf Asymptotic structure of the Pauli-Fierz theory in four spacetime dimensions}

\vspace{30pt}
{Oscar Fuentealba${}^{\, a}$, Marc Henneaux${}^{\, a,b}$,  Sucheta Majumdar${}^{\, a}$, Javier Matulich${}^{\, a}$ \\ and C\'edric Troessaert${}^{\, a,c}$}

\vspace{10pt}
\texttt{oscar.fuentealba@ulb.ac.be,henneaux@ulb.ac.be,sucheta.majumdar@ulb.ac.be, jmatulic@ulb.ac.be,cedric.troessaert@hers.be}

\vspace{20pt}
\begin{enumerate}
\item[${}^a$] {\sl \small
Universit\'e Libre de Bruxelles and International Solvay Institutes,\\
ULB-Campus Plaine CP231, B-1050 Brussels, Belgium}
\item[${}^b$] {\sl \small
Coll\`ege de France, 11 place Marcelin Berthelot, 75005 Paris, France\\}
\item[${}^c$] {\sl \small
Haute-Ecole Robert Schuman, Rue Fontaine aux M\^ures, 13b, B-6800, Belgium}
\end{enumerate}

\vspace{50pt}
{\bf Abstract} 
\end{center}

\noindent
The asymptotic structure of the Pauli-Fierz theory at spatial infinity  is investigated in four spacetime dimensions.  Boundary conditions on the massless spin-$2$ field that are invariant under an infinite-dimensional group of non-trivial ``improper'' gauge symmetries are given. The compatibility of these boundary conditions with invariance of the theory under Lorentz boosts is a subtle issue which is investigated in depth and leads  to the identification of the improper gauge symmetries with the pure BMS supertranslations. It is also shown how rigid Poincar\'e transformations and improper gauge symmetries of the free Pauli-Fierz theory merge into the full BMS group as one switches on the gravitational coupling.  Contrary to the massless spin-$1$ case, where invariance under boosts is implemented differently and where important differences between the free and the interacting cases have been exhibited recently, the free Pauli-Fierz theory and general relativity show very similar behaviours at spatial infinity.

\newpage

\setcounter{tocdepth}{2}
\tableofcontents

\newpage

\section{Introduction}
\setcounter{equation}{0}

The study of the asymptotic properties of gravity in the asymptotically flat context is a remarkably rich subject that has undergone a revived interest in the last years \cite{Strominger:2017zoo}.  The group of asymptotic symmetries (analysed either at null infinity \cite{Bondi:1962px,Sachs:1962wk,Sachs:1962zza,Penrose:1962ij,Madler:2016xju,Alessio:2017lps,Ashtekar:2018lor} or at spatial infinity \cite{Geroch:1972up,Ashtekar:1978zz,Compere:2011ve,Troessaert:2017jcm,Henneaux:2018cst,Henneaux:2018hdj,Henneaux:2019yax}) is infinite-dimensional and called the ``BMS group''.  It contains, besides the anticipated Poincar\'e transformations, an infinite number of ``supertranslations'' parametrized by one arbitrary function of the angles (coordinates on the $2$-sphere).   While the Poincar\'e transformations are just the isometries of the background Minkowski space to which the geometry is asymptotic at infinity, the meaning of the pure supertranslations (i.e., of the supertranslations that are not standard translations) is more subtle and has been brought later in terms of geometric constructions at null infinity \cite{Penrose74,Schmidt77,Geroch77}.

In order to shed further light on the emergence of the BMS group and the nature of the supertranslations, we analyse in this paper the asymptotic symmetries of the free massless spin-$2$ theory (Pauli-Fierz theory \cite{Fierz:1939ix}).  The analysis is carried out at spatial infinity and uses concepts that are directly available there.

It turns out -- and we want to stress it from the outset -- that Poincar\'e invariance (more precisely, invariance under Lorentz boosts) plays a central role in the analysis.  Indeed, as we shall show,  compatibility of a symplectic action of the Poincar\'e group with boundary conditions at spatial infinity that leads to a non-trivial infinite-dimensional asymptotic symmetry group, is a subtle question that has an important impact on the asymptotic structure.  

Given the importance of Poincar\'e invariance, we now give a brief overview of where it comes into play.  The discussion also exhibits somewhat unanticipated differences between the free massless spin-$1$ and spin-$2$ fields.

\subsubsection*{Pure supertranslations}
 
The Pauli-Fierz theory is a Lorentz-invariant theory for a symmetric tensor $h_{\mu \nu}$, formulated in Minkowski space.  It possesses an abelian gauge symmetry parametrized by a vector field. From that point of view, it appears to be a direct generalization of the free Maxwell theory, differing only in the tensorial nature of the dynamical fields and of the gauge parameters.  In spite of their formal similarities, we shall see that there is, however, an important difference between the free spin-$1$ and free spin-$2$ massless theories in the way relativistic invariance is made compatible with the asymptotic structure. This is the major dissimilarity between the two dynamical systems that we alluded to above. The different implementation of Lorentz invariance for the spin-$2$ theory leads to  a reduction of the size of the group of ``improper'' (sometimes called ``global'' or ``large'')  gauge transformations of the Pauli-Fierz theory with respect to what a naive generalization from spin-$1$ to spin-$2$ would wrongly suggest.

To be more specific, we recall that among the gauge symmetries of a gauge theory, those that do not have a vanishing charge (given by a non-trivial surface integral) do change the physical state of the system.  They were called ``improper gauge transformations'' in the article \cite{Benguria:1976in}, which gave illuminating insights into that question, and we shall adopt that terminology here. By contrast, the gauge transformations with a generator that vanishes for all field configurations correspond to redundancies that must be factored out.  They are called ``proper''.

The gauge symmetries of the vector potential of the Maxwell theory take the form
\be
\delta_{\epsilon} A_\mu =  \partial_\mu \epsilon  \label{eq:GaugeSymmSpin1}
\ee
where the gauge parameter $\epsilon$ is a scalar.  Among these, some are proper and some are improper.  The distinction between the two can be made only after precise boundary conditions on the fields have been given.  This was achieved in \cite{Henneaux:2018gfi}, where it was shown that the improper gauge transformations were characterized by a single function of the angles on the sphere at infinity that can be connected -- although in a somewhat subtle way -- with the asymptotic form of the single gauge parameter $\epsilon$   (see also \cite{Balachandran:2013wsa,Campiglia:2017mua} for studies of the asymptotics of the electromagnetic field at spatial infinity).  The improper gauge symmetry group of the spin-$1$ theory  is therefore described by ``angle-dependent $u(1)$ transformations'', in complete agreement with the null infinity findings \cite{Strominger:2013lka,Barnich:2013sxa,He:2014cra,Lysov:2014csa,Kapec:2014zla,Kapec:2015ena,Campiglia:2016hvg,Conde:2016csj}. 

The gauge symmetries of the free spin-$2$ theory are
\be
\delta_{\epsilon} h_{\mu \nu} =  \partial_\mu \epsilon_\nu + \partial_\nu \epsilon_\mu \, ,   \label{eq:GaugeSymmSpin2}
\ee
where $\epsilon_\mu$ is now a vector and is accordingly described by four independent spacetime functions.
The great similarity between the symmetry transformations (\ref{eq:GaugeSymmSpin1}) of the free massless spin-1 theory and (\ref{eq:GaugeSymmSpin2}) of the free massless spin-2 theory might lead one to expect that the asymptotic symmetries of the Pauli-Fierz theory are just the fourfold of those of the Maxwell theory,  the  improper gauge transformations being parametrized by the asymptotic values of $\epsilon_\mu$.  As we show in this paper, it turns out that this is not correct, but due to asymptotic subtleties, the improper gauge transformations of the Pauli-Fierz theory are parametrized by a single function of the angles, and not four.

The reduction from four functions of the angles to one function of the angles is the result of asymptotic conditions  that relate the asymptotic form of the allowed angular gauge transformations to the asymptotic form of the radial gauge transformations.   These reduction conditions,  which have no analog in electromagnetism, are crucial because they naturally implement the request that the Lorentz boosts should have a canonical action, guaranteeing a well-defined moment map for the full Poincar\'e group, which appears as a rigid symmetry group in the linearized spin-$2$ theory.  They are also somewhat reminiscent to conditions arising in the asymptotically AdS context \cite{Henneaux:1985tv,Brown:1986nw}.

We recall that a ``symmetry'' of a theory is more than just a transformation leaving the boundary conditions invariant.  It should also leave the action invariant. Thus, by ``Poincar\'e invariance'', we mean not only invariance of the boundary conditions under Poincar\'e transformations but, equally crucially, invariance of the Pauli-Fierz action itself up to boundary terms at the initial and final times but {\bf not} at spatial infinity. Invariance of the action guarantees that the transformations are canonical since they will automatically leave the symplectic form strictly invariant (and not just invariant up to a surface term). The relativistic invariance of the action up to boundary terms is guaranteed by the Lorentz covariance of the Pauli-Fierz Lagrangian density, but the vanishing of the boundary terms at spatial infinity (for all configurations fulfilling the given boundary conditions) turns out to be a non-trivial requirement.

\subsubsection*{Turning on the gravitational coupling}

The fact that the improper gauge symmetries of the free spin-$2$ theory match in number the BMS supertranslations of the full Einstein theory enables one  to view these BMS supertranslations as the ``lifts'' to the nonlinear Einstein theory of the improper  gauge transformations of the Pauli-Fierz theory.   This is a non-trivial property since in the Maxwell case, the direct lift at spatial infinity of the infinite-dimensional symmetry algebra of angle-dependent $u(1)$ transformations to the nonlinear Yang-Mills case is obstructed due to difficulties with the boosts \cite{Tanzi:2020fmt}. The asymptotic symmetries can therefore be affected by the interactions, and this opens the logical possibility that the free spin-$2$ theory could have a larger asymptotic symmetry group than general relativity.  Our results show that this is not the case. 

As one switches on the interactions in the spin-$2$ theory,  the improper gauge symmetries of the free theory merge with the rigid Poincar\'e symmetries to form the entire BMS group (which is composed only of improper diffeomorphisms in the full Einstein theory).  We also clarify in this paper how this is realized. 
In particular, we indicate why  the asymptotic translations are {\bf not} counted twice, once as emerging from the rigid translation subgroup of the Poincar\'e group and once as emerging from the improper gauge symmetries of the free theory with constant coefficients. The answer to this question has interesting group theoretical roots and is  given in our analysis. 

Boosts thus play  a crucial role in the structure of the asymptotic symmetries.  We have found that Einstein's theory behaves in that respect as the free Pauli-Fierz theory at spatial infinity, in contrast to the massless spin-$1$ situation.  It is not inappropriate to recall in that context that the ``boost problem'' takes a very different form in general relativity and in Yang-Mills theory \cite{Christodoulou1981}, where it was also found that Einstein's theory behaves for that question as a free theory.

\subsubsection*{Organization of paper}

Our paper is organized as follows.  In the next section, we recall the action and the gauge symmetries of the free Pauli-Fierz theory.  We then formulate in Section \ref{Asymptotia} the boundary conditions, paying due attention to finiteness of the action. Section \ref{sec:CanoniBoosts} is devoted to the conditions resulting from the requirement that the boosts should have a symplectic action and works out their main consequences.  The electromagnetic situation is compared and contrasted. In Section \ref{sec/CWithinL}, we derive the charge-generators of the symmetries and show in particular that the boost and rotation generators must be supplemented by non-trivial surface integrals in addition to their standard non vanishing bulk piece. The algebra of the charges, and how pure supertranslations and translations algebraically fit together, are questions discussed in Section \ref{sec:Algebra}.   Section  \ref{sec:WeakField_Charges} rederives the charges of the free Pauli-Fierz theory from the weak field expansion of the charges of the Einstein theory.  Section \ref{sec:Conclusions} contains conclusions and comments.  Finally, four appendices of a more technical nature complete our paper.

\section{Action and gauge symmetries}

\subsection{Action}

We start with the action of linearized gravity on Minkowski spacetime in Hamiltonian form, which
reads, 
\begin{equation}
I[h_{ij}, \pi^{ij}, n, n^i] =\int dtd^{3}x\left(\pi^{ij}\dot{h}_{ij}-\mathcal{E}-n\mathcal{G}-n^{i}\mathcal{G}_{i}\right)\,.\label{eq:Action}
\end{equation}
Here, the dynamical fields are the conjugate pairs $(\pi^{ij},h_{ij})$,
while $n$ and $n^{i}$ are the Lagrange multipliers associated
to the following constraints 
\begin{align}
\mathcal{G} & =-\sqrt{\gamma}(\nabla^{i}\nabla^{j}h_{ij}-\triangle h)\,,\label{eq:HamG}\\
\mathcal{G}_{i} & =-2\nabla^{j}\pi_{ij}\,.\label{eq:MomG}
\end{align}
The symbol $\nabla_{i}$ denotes the covariant derivative with respect to the flat Euclidean metric $\gamma_{ij} $ ($= \delta_{ij}$ in cartesian coordinates) and $\triangle \equiv \nabla ^i \nabla_i$.  Indices are lowered and raised with the flat metric $\gamma_{ij}$.   Lastly, $\gamma$ is its determinant.  While we shall use cartesian coordinates where $\nabla_{i} = \partial_i$ and $\gamma = 1$, we shall also find it convenient to work in polar coordinates where the Christoffel symbols do not vanish.

The energy and momentum densities are given by 
\begin{eqnarray}
\mathcal{E} & = & \frac{1}{\sqrt{\gamma}}\left(\pi^{ij}\pi_{ij}-\frac{\pi^{2}}{2}\right)+\sqrt{\gamma}\left(\frac{1}{4}\nabla_{k}h_{ij}\nabla^{k}h^{ij}-\frac{1}{2}\nabla_{j}h^{ij}\nabla^{k}h_{ik}+\frac{1}{4}\nabla_{i}h\nabla^{i}h\right) \nonumber \\
 &  & +\sqrt{\gamma}\nabla_{l}\left(-h^{ij}\nabla^{l}h_{ij}-h^{il}\nabla_{i}h+\frac{3}{2}h^{lj}\nabla^{i}h_{ij}+\frac{1}{2}h_{ij}\nabla^{i}h^{jl}\right)+\frac{1}{2}h\mathcal{G}\,, \label{eq:EnerDen}\\
\mathcal{P}_{i} & = & -2\partial_{j}\left(\pi^{jk}h_{ik}\right)+\pi^{jk}\partial_{i}h_{jk}\,, 
\end{eqnarray}
respectively.   In (\ref{eq:EnerDen}), one may drop the last term if one so wishes, proportional to the constraints.

The action (\ref{eq:Action}) follows from the canonical action for gravity \cite{Dirac:1958jc,Arnowitt:1962hi} upon linearization.  The derivation is recalled in  Appendix \ref{App:Linear}.

\subsection{Symmetries}

The Einstein-Hilbert action from which the free massless spin-$2$ theory descends is invariant under diffeomorphisms. This endows the Pauli-Fierz action with both rigid and gauge symmetries (see Appendix \ref{App:Linear}).

\subsubsection{Covariant expression}

The rigid symmetries are the background symmetries and take the form
\be
\delta_\xi h_{\mu \nu} = \mathcal{L}_\xi h_{\mu \nu}\, ,   
\ee
where the Lie derivative of $h_{\mu \nu}$ is given by
\be
\mathcal{L}_\xi h_{\mu \nu} = \xi^\rho \partial_\rho h_{\mu \nu} + \partial_\mu \xi^\rho h_{\rho \nu} + \partial_\nu \xi^\rho h_{\mu \rho } 
\ee
and where $ \xi^\mu$ is a Killing vector of the Minkowskian metric ($\mathcal{L}_\xi g_{\mu \nu}^{\textrm{Minkowski}}= 0$) and thus an infinitesimal Poincar\'e transformation.
The gauge transformations (proper and improper) are the infinitesimal diffeomorphisms of the background metric and read 
\be
\delta_\epsilon h_{\mu \nu} =  \mathcal{L}_\epsilon g_{\mu \nu}^{\textrm{Minkowski}}  = \nabla_\mu \epsilon_\nu + \nabla_\nu \epsilon_\mu \, , \label{eq:Gauge0}
\ee
where $\epsilon^\mu$ is an arbitrary vector field, submitted only to asymptotic conditions that will be made precise below.

The Poincar\'e transformations are linear in the fields $h_{\mu \nu}$ while the gauge transformations do not depend on them.  Putting the two together, one has
\be
\delta_{\xi, \epsilon} h_{\mu \nu} = \mathcal{L}_\xi h_{\mu \nu} + \nabla_\mu \epsilon_\nu + \nabla_\nu \epsilon_\mu \, ,   \label{eq:SymmSpin2}
\ee
which reduces to 
\be
\delta_{\xi, \epsilon} h_{\mu \nu} = \mathcal{L}_\xi h_{\mu \nu} + \partial_\mu \epsilon_\nu + \partial_\nu \epsilon_\mu   \label{eq:SymmSpin2VM}
\ee
in Minkowskian coordinates.

It is clear from (\ref{eq:Gauge0}) that if $\epsilon^\mu$ itself is a Killing vector of the Minkowski metric, it has no action on the field $h_{\mu \nu}$.  The symmetry transformations are thus redundant and we can factor the space of the $\epsilon^\mu$'s by the Poincar\'e transformations (already contained in the $\xi^\mu$'s) to avoid the redundancy. 

It is also clear that we have the freedom to redefine the rigid symmetries $\mathcal{L}_\xi h_{\mu \nu}$  by adding to them a gauge symmetry $\delta_{\epsilon(\xi)} h_{\mu \nu}$, where the gauge symmetry parameter $\epsilon_\mu(\xi)$ is some definite function of the Poincar\'e Killing vectors.  This freedom will turn out to be crucial when analysing the integrability of the boost generators.

To discuss the dynamical implementation of the symmetries and the corresponding moment map, it is convenient to revert to the Hamiltonian formulation. 
Our first task is thus to rewrite the above symmetries in Hamiltonian form. 

\subsubsection{Poincar\'e transformations in Hamiltonian form}
We first consider the rigid Poincar\'e symmetries.
The transformations of the canonical conjugate pairs can be obtained from the variations of the Lagrangian variables of the free theory through standard methods.   Alternatively, one can just linearize the known Poincar\'e transformations rules in the full Einstein theory.  Either way, one gets, with $(\xi^\mu) \rightarrow (\xi^\perp\equiv \xi, \xi^i)$, 
\begin{eqnarray}
\delta_{\xi}h_{ij} & = & \frac{2\xi}{\sqrt{\gamma}}\left(\pi_{ij}-\frac{1}{2}\gamma_{ij}\pi\right)+\mathcal{L}_{\xi}h_{ij}\,,\label{dh-Poincare}\\
\delta_{\xi}\pi^{ij} & = & \frac{1}{2}\sqrt{\gamma}\xi \left(\triangle h^{ij}+\nabla^{i}\nabla^{j}h-2\nabla^{(i}\nabla_{k}h^{j)k}\right)\nonumber \\ && +\frac{1}{2}\sqrt{\gamma}\nabla_{k}\xi \left[\nabla^{k}h^{ij}-2\nabla^{(i}h^{j)k}+\gamma^{ij}\left(2\nabla_{l}h^{kl}-\nabla^{k}h\right)\right]\nonumber \\
 &  & +\sqrt{\gamma}\triangle\xi \, h^{ij}+\sqrt{\gamma}\gamma^{ij}\nabla_{k}\nabla_{l}\xi\,  h^{kl}-2\sqrt{\gamma}\nabla_{k}\nabla^{(i}\xi \, h^{j)k} \nonumber \\ && - \frac{1}{2}\sqrt{\gamma}\left(\nabla^{i}\nabla^{j}\xi-\gamma^{ij}\triangle \xi \right)h - \frac{1}{2}\gamma^{ij}\xi \, \mathcal{G}+\mathcal{L}_{\xi}\pi^{ij}\,.\label{eq:dp-Poincare}
\end{eqnarray}
The first two terms in the last line of (\ref{eq:dp-Poincare}) come from the $(-1/2) h \mathcal{G}$ term in $\mathcal{E}$.  Furthermore,  $\mathcal{L}_{\xi}h_{ij} $ and $\mathcal{L}_{\xi}\pi^{ij}$ are the spatial Lie derivatives,
\begin{align}
\mathcal{L}_{\xi}h_{ij} & =2h_{k(i}\partial_{j)}\xi^{k}+\xi^{k}\partial_{k}h_{ij}\,,\\
\mathcal{L}_{\xi}\pi^{ij} & =-2\partial_{k}\xi^{(i}\pi^{j)k}+\partial_{k}\left(\xi^{k}\pi^{ij}\right)\,.
\end{align}
In Minkowskian coordinates,  the generators $\xi$ and $\xi^i$ of the Poincar\'e group take the form
\be
\xi = b_i x^i + a^0, \qquad \xi^i = {b^i}_j x^j + a^i, \label{eq:KVMinkowski}
\ee
where $b_i$, $b_{ij} = - b_{ji}$, $a^0$ and $a^i$ are constants (we consider fixed time slices, and at any given time, one may absorb the term $b^i x^0$ in $\xi^i$ in the space translation $a^i$).

As we pointed out in the previous section, there is some ambiguity in the form of the Poincar\'e transformations, to which one can add some definite gauge transformation. 

\subsubsection{Gauge symmetries in Hamiltonian form}
The gauge symmetry transformations read, in Hamiltonian form,
\begin{eqnarray}
\delta_{\epsilon}h_{ij} & = & \nabla_{i}\epsilon_{j}+\nabla_{j}\epsilon_{i}\,,\label{eq:dh-Gauge}\\
\delta_{\epsilon}\pi^{ij} & = & \sqrt{\gamma}\left(\nabla^{i}\nabla^{j}\epsilon-\gamma^{ij}\triangle\epsilon \right)\,.\label{eq:dp-Gauge}
\end{eqnarray}
It is again clear that if we take for $(\epsilon, \epsilon_i)$ the Killing vectors (\ref{eq:KVMinkowski}) of Minkowski space, one gets identically zero. 

As shown below, the boundary conditions on $h_{ij}$ and $\pi^{ij}$ contain in particular the requirements that $h_{ij} = O(\frac{1}{r})$ and $\pi^{ij} = O(\frac{1}{r^2})$.  The gauge transformations (\ref{eq:dh-Gauge}) that preserve the condition $h_{ij} = O(\frac{1}{r})$ must fulfill $ \partial_i \epsilon_j + \partial_j \epsilon_i = O(\frac{1}{r})$, from which one gets $\partial_i \partial_j \epsilon_k =  O(\frac{1}{r^2})$.   A first integration yields $\partial_j \epsilon_k =  b_{jk} + O(\frac{1}{r})$, where $b_{jk}$ are constants (see Appendix \ref{App0}).  A second integration implies then
\be
\epsilon_k = b_{kj} x^j + a_k \ln r + f_k(\mathbf{n}) + O\left(\frac{1}{r} \right) \, ,
\ee
where $b_{kj}$ is antisymmetric, $b_{kj} = -b_{jk}$, in order to fulfill the original equation, $a_k$ are constants and  $f_k(\mathbf{n})$ are functions of the angles only (see again Appendix \ref{App0}; $\mathbf{n}$ is the unit normal to the spheres centered at the origin, $n^i = \frac{x^i}{r}$, and depends only on the angles on the spheres).

The logarithmic term yields the contribution 
$ \frac{1}{r} (a_i n_j + a_j n_i)$ to $h_{ij}$ and hence $\frac{2}{r} \mathbf{a} \cdot \mathbf{n}$ to $h_{rr}$.  But there are extra boundary conditions that imply that the leading order of $h_{rr}$ should be even under the parity transformation $\mathbf{n} \rightarrow -\mathbf{n}$ (see Eq. (\ref{eq:AsympLambda}) below), forcing the constant vector $a_i$ to be zero\footnote{The logarithmic term defines ``logarithmic'' translations  \cite{Bergmann:1961zz,AshtekarLog1985} and is part of a wider class of transformations, the ``logarithmic supertranslations'' where $a_k$ is allowed to depend on the angles.  We are currently carrying the task of devising consistent, more flexible,  boundary conditions which incorporate these logarithmic transformations as (improper) gauge symmetries \cite{FHMMT}.}.
Furthermore, $b_{kj} x^j$  defines an isometry of Euclidean space (rotation) and so can be discarded since it yields no variation of the fields.  This means that without loss of generality we can assume
\be
\epsilon_k = \xbar \epsilon_k(\mathbf{n})+ O\left(\frac{1}{r}\right) \label{eq:ParameterGaugeTransf}
\ee
and this is what we shall do from now on.

There is still some redundancy in the description since the spatial translations are contained in (\ref{eq:ParameterGaugeTransf}).  A consistency condition, which will be verified to hold below,  is that they should yield a zero charge (as gauge transformations, not as part of the rigid Poincar\'e symmetry group).

A similar reasoning, using the asymptotic form of $\pi^{ij}$ given below,  shows that  one can take
\be
\epsilon = \xbar \epsilon(\mathbf{n})+ O\left(\frac{1}{r}\right) \label{eq:ParameterGaugeTransfTime}
\ee
and this is the form of $\epsilon$ that we shall adopt from now on.  Again, there is still a redundancy, given by the time translations (zero mode of $f$).

As we shall show in Section \ref{sec:CanoniBoosts} below, the functions $\xbar \epsilon_k(\mathbf{n})$ will be subject to further conditions, given by (\ref{eq:FormEpsilonK}).

\section{Asymptotic conditions and symplectic structure \label{Asymptotia}}

The boundary conditions on the conjugate pairs $(h_{ij}, \pi^{ij})$ define the allowed phase space configurations to be included in the theory and complete thereby the definition of phase space.  We present in this paper boundary conditions that make the canonical action ``$\int dt (p \dot{q} - H)$'' finite off-shell (and not just on-shell) without need for regularization.  This enables one to use standard Hamiltonian methods and, in particular to construct appropriate moment maps, without having to deal with infinities, the removal of which might involve ambiguities. 

As in the Maxwell theory and in the Einstein theory, the boundary conditions on $h_{ij}$ and $\pi^{ij}$ involve three ingredients \cite{Henneaux:2018hdj,Henneaux:2019yax}:
\begin{enumerate}
\item fall-off conditions;
\item ``gauge-twisted''  parity conditions on the leading terms in the asymptotic expansion, which generalize the strict parity conditions of \cite{Regge:1974zd} in order to accommodate the BMS symmetry;
\item (technical:) stronger fall-off of the constraints than the one dictated by the fall-off of the fields.
\end{enumerate}
A fourth key ingredient will be needed below (Section \ref{sec:CanoniBoosts}), namely, a stronger fall-off of the mixed radial-angular components of $h_{ij}$, but we do not impose it at first as we want to stress how the need for it arises.   This fourth ingredient is present in the complete Einstein theory but has no direct analog in the Maxwell case.

\subsection{Cartesian coordinates}

The fall-off of the spin-$2$ field and its conjugate momentum is the one characteristic of massless fields and reads
\begin{eqnarray}
h_{ij} &=& \frac{\xbar h_{ij}(\mathbf n)}{r} + \frac{ h^{(2)}_{ij}(\mathbf n)}{r^2} + O\left(\frac{1}{r^3} \right) \, , \label{eq:DecayFields0}\\
\pi^{ij} &=& \frac{\xbar \pi^{ij}(\mathbf n)}{r^2} + \frac{ \pi_{(2)}^{ij}(\mathbf n)}{r^3} + O\left(\frac{1}{r^4} \right)\, .
\label{eq:DecayFields1}
\end{eqnarray}

These conditions are supplemented by gauge-twisted parity conditions.
In order to formulate them, we decompose the leading terms  into even and odd parts under the parity transformation $\mathbf n \rightarrow -\mathbf n$,
\begin{equation}
\xbar{h}_{ij} = (\xbar h_{ij})^{\text{even}} + (\xbar{h}_{ij})^{\text{odd}}\ .
\end{equation}
We do not assume that $\xbar h_{ij}$  is strictly even, i.e., that $(\xbar h_{ij})^{\text{odd}}$ vanishes.   Rather, we allow here for a ``twist'' by a non-vanishing $(\xbar h_{ij})^{\text{odd}}$ which is imposed to take the form of a gauge transformation, 
\be
(\xbar h_{ij})^{\text{odd}}= \nabla_{i}\zeta_{j}\ +\ \nabla_{j}\zeta_{i}\ ,\label{TPh}
\ee
while 
$(\xbar h_{ij})^{\text{even}}$ remains arbitrary.  The vector $\zeta_{j}$ describes a gauge transformation and so must asymptotically behave as in (\ref{eq:ParameterGaugeTransf}).  Furthermore, we can assume, as we have done,  that it reduces to its leading $O(r^0)$ part $\xbar \zeta_{j}$ (subleading terms will modify only the unrestricted subleading components of $h_{ij}$), and that this leading part $\xbar \zeta_{j}$ is even, since the contribution of the odd part can be absorbed in a redefinition of $(\xbar h_{ij})^{\text{even}}$.  The form of $(\xbar \zeta_{i})^{\text{even}}$ will be found in Section \ref{sec:CanoniBoosts} to need further restrictions (specifically through (\ref{eq:FormZeta})), but again, at this stage, we keep it unrestricted, to show how these further conditions arise.

Similarly, for the conjugate
momenta $\bar{\pi}^{ij}$ we assume the following twisted parity condition
\begin{equation}
\xbar{\pi}_{ij} = (\xbar{\pi}_{ij})^{\text{odd}} + (\xbar \pi_{ij})^{\text{even}} \, ,
\label{TPp}
\end{equation}
where the odd component $(\xbar{\pi}_{ij})^{\text{odd}}$ is arbitrary and where $(\xbar \pi_{ij})^{\text{even}}$ takes the form of a gauge transformation,
\begin{equation}
(\xbar \pi_{ij})^{\text{even}} =\nabla^{i}\nabla^{j}V\ -\ \delta^{ij}\triangle V \ .\label{TPp2}
\end{equation}
The $O(r^0)$ function $V(\mathbf{n})$
is assumed to be even. 

The constraints \eqref{eq:HamG}, \eqref{eq:MomG} are linear in the fields and homogeneous in derivatives.  Thus, they split into independent equations for each order in the asymptotic expansion.  When the fields obey the decay (\ref{eq:DecayFields0})-(\ref{eq:DecayFields1}), the constraint functions  $\mathcal{G}$ and ${\mathcal G}_i$ are of order $O(1/r^3)$.  The third ingredient in the asymptotic boundary conditions is that the only allowed off-shell configurations should be such that $\mathcal{G}$ and ${\mathcal G}_i$ are of order $O(1/r^4)$ (the leading term and the subleading term in the constraint equations are fulfilled, but we allow off-shell configurations that may not obey the subsequent lowest order of the constraints).

We therefore impose
\begin{equation}
\left(\nabla^{i}\nabla^{j}-\delta^{ij} \triangle\right)\left(\frac{\xbar h_{ij}(\mathbf n)}{r}\right)=0\,,\qquad \nabla_{i}\left(\frac{\xbar \pi^{ij}(\mathbf n)}{r^2}\right)=0\ .\label{odd-even const0}
\end{equation}
Because the improper gauge components satisfy
the constraints, which are invariant under proper and improper gauge transformations,  Eqn. (\ref{odd-even const0}) reduces to
\begin{equation}
\left(\nabla^{i}\nabla^{j}-\delta^{ij} \triangle\right)\left(\frac{(\xbar h_{ij})^{\text{even}}(\mathbf n)}{r}\right)=0\,,\qquad\nabla_{i}\left(\frac{ (\xbar{\pi}_{ij})^{\text{odd}}(\mathbf n)}{r^2}\right)=0\, .\label{odd-even const}
\end{equation}
There is  no contribution of
$\zeta_{i}$ and $V$ in \eqref{odd-even const}.

The boundary conditions can easily be checked to be Poincar\'e invariant.  This is because the boosts and spatial rotations have a definite odd parity and furthermore, pure gauge (proper or improper) variations of the variables remain pure gauge.  In addition, the constraints are first class and preserved order by order under Poincar\'e transformations. 

The boundary conditions are also invariant under the gauge transformations, provided the gauge parameters behave as in (\ref{eq:ParameterGaugeTransf}) and (\ref{eq:ParameterGaugeTransfTime}).

\subsection{Finiteness of the action}
Any phase space history $(h_{ij}(t,x^k), \pi^{ij}(t, x^k))$ obeying the above  boundary conditions make the action finite, independently as to whether it fulfills the equations of motion.  

To see this, we note that there are two terms in the action, \ie, the ``kinetic term $\int dt p \dot{q}$'', which defines the symplectic structure, and the term involving the Hamiltonian.   Since the energy density $\mathcal{E}$ decays as $1/r^4$ at infinity, this second term is manifestly finite.  

We only need to consider the kinetic term.   The reason that parity conditions are imposed on top of the decay (\ref{eq:DecayFields0})-(\ref{eq:DecayFields1}) is actually precisely the need to make the kinetic term in the action and the corresponding symplectic form finite.  They would otherwise be plagued by logarithmic divergences.  We impose twisted parity conditions as opposed to strict parity conditions, because strict parity conditions might require an improper gauge fixing and hence are not always available. This is exactly as in the full nonlinear theory, and the verification that the kinetic term is finite proceeds exactly as in \cite{Henneaux:2018hdj,Henneaux:2019yax}.  This verification uses the fact that the leading term of the constraint functions is zero.

It is not a surprise that the proof of finiteness of the kinetic term proceeds as in the full Einstein theory, since the kinetic term takes the same form and the boundary conditions at infinity are the same.  It is a rather generic reasoning, which is also valid for electromagnetism \cite{Henneaux:2018gfi}.  The differences with the Maxwell theory appear at a later stage.

Another way to deal with the logarithmic divergence would be to impose no parity conditions at all but subtract the generically divergent term $\lim_{r \rightarrow \infty} \log r \int dt  d\theta  d \varphi \xbar \pi^{ij}  \dot{\xbar h}_{ij}$ 
to regulate the action and the symplectic form \cite{Compere:2011ve}.  This is not the path followed here since it is not necessary to allow a more general asymptotic behaviour to get an interesting asymptotic structure.  Our procedure is manifestly finite throughout, consistently avoiding infinities and enabling one to use the methods of finite-dimensional Hamiltonian mechanics without having to bring in regularizations. In particular, if the phase space vector field defined by the evolution equations leaves the symplectic form invariant -- strictly and not just up to a surface term --,  it has a well-defined generator (the Hamiltonian!) and the classical solutions of the equations (with suitable boundary conditions at initial and final times) are true stationary points of the action  -- and not just stationary points up to surface terms at spatial infinity.

To avoid a possible source of confusion, we end this section with a comment on our conventions for evaluating the asymptotic behaviour of integrals involving fields expressed in cartesian coordinates, e.g., $\int d^3x \pi^{ij} \dot{h}_{ij}$.  We first compute $\pi^{ij} \dot{h}_{ij}$ in cartesian coordinates, and then go to spherical coordinates $(x^i) \rightarrow (r, \theta, \varphi)$.  The Jacobian brings in $r^2 \sin \theta$, $(d^3 x)_{\textrm{cartesian}} = r^2 \sin \theta dr d \theta d \varphi$.  Of course, one gets the same result when dealing directly with fields expressed in spherical coordinates, since the integral of a density of weight one is invariant.  The factor $ r^2 \sin \theta$ does not originate then from any Jacobian since $(d^3 x)_{\textrm{spherical}} = dr d \theta d \varphi$ but from the field $\pi^{ij}$ (or any other similar field), which carries a unit density weight.  The context should always make the derivation clear.

\subsection{Spherical coordinates}

\subsubsection{Asymptotic conditions}

The flat metric $- dt^2 + \delta_{ij} dx^i dx^j$ reads, in spherical coordinates,
\be
ds^2 = - dt^2 + dr^2 + \gamma_{AB} dx^A dx^B \, ,
\ee
where 
\be
\gamma_{AB} = r^2 \xbar \gamma_{AB} \ .
\ee
Here, $\xbar \gamma_{AB}$ is the metric on the round unit sphere, and $x^A$ are coordinates on the sphere which we will call ``the angles''.  With the traditional variables $\theta$ and $\varphi$ one has $\xbar \gamma_{AB} dx^A dx^B = d \theta^2 + \sin^2 \theta d \varphi^2$.   We denote the covariant derivative on the unit sphere by $\xbar D_A$ and we set $\xbar \triangle \equiv \xbar D_A \xbar D_B \xbar \gamma^{AB} \equiv \xbar D_A \xbar D^A$ (barred quantities live on the unit sphere, and we lower and raise their indices with the metric $\xbar \gamma_{AB}$ and its inverse $\xbar \gamma^{AB}$). 

The form of the antipodal map depends on the coordinates being used on the sphere. With $\theta$ and $\varphi$, it is $\theta \rightarrow \pi - \theta$ and $\varphi \rightarrow \varphi + \pi$.  We shall assume for definiteness that we have chosen the ``angles'' $x^A$ such that the antipodal map reads $x^A \rightarrow - x^A$. 

In spherical coordinates, the above fall-off of the metric components is
\begin{eqnarray}
h_{rr} & = & \frac{1}{r}\bar{h}_{rr}+\frac{1}{r^{2}}h_{rr}^{(2)}+\mathcal{O}(r^{-3})\,,\\
h_{rA} & = & \bar{\lambda}_{A}+\frac{1}{r}h_{rA}^{(2)}+\mathcal{O}(r^{-2})\,,\\
h_{AB} & = & r\bar{h}_{AB}+h_{AB}^{(2)}+\mathcal{O}(r^{-1})\,,
\end{eqnarray}
while the asymptotic conditions on the momentum are given by 
\begin{eqnarray}
\pi^{rr} & = & \bar{\pi}^{rr}+\frac{1}{r}\pi_{(2)}^{rr}+\mathcal{O}(r^{-2})\,,\\
\pi^{rA} & = & \frac{1}{r}\bar{\pi}^{rA}+\frac{1}{r^{2}}\pi_{(2)}^{rA}+\mathcal{O}(r^{-3})\,,\\
\pi^{AB} & = & \frac{1}{r^{2}}\bar{\pi}^{AB}+\frac{1}{r^{3}}\pi_{(2)}^{AB}+\mathcal{O}(r^{-4})\,.
\end{eqnarray}
The parity conditions under the
antipodal map on the leading order components of the spin-$2$ field and its conjugate momentum read in polar coordinates,
\begin{eqnarray}
 & \xbar{h}_{rr}=\text{even}\,,\quad \xbar{\lambda}_{A}=(\xbar{\lambda}_{A})^{\text{odd}} + \xbar D_A \zeta_r - \xbar \zeta_A \,, \label{eq:AsympLambda}\\
 & \xbar{\pi}^{rr}=(\xbar{\pi}^{rr})^{\text{odd}}\ -\ \sqrt{\xbar{\gamma}}\xbar{\triangle}V\,,\quad\xbar{\pi}^{rA}=(\xbar{\pi}^{rA})^{\text{even}}-\sqrt{\xbar{\gamma}}\xbar{D}^{A}V\,, \\
 & \xbar{\pi}^{AB}=(\xbar{\pi}^{AB})^{\text{odd}}+\sqrt{\xbar{\gamma}}(\xbar{D}^{A}\xbar{D}^{B}V-\xbar{\gamma}^{AB}\xbar{\triangle}V)\,, \\
 & \xbar{h}_{AB}=(\xbar{h}_{AB})^{\text{even}}+\xbar{D}_{A}\xbar \zeta_{B}+ \xbar{D}_{B}\xbar \zeta_{A}+2\xbar{\gamma}_{AB}\zeta_r\,. \label{eq:AsympHAB}
\end{eqnarray}
One has $\zeta_i dx^i = \zeta_r dr + \zeta_A dx^A$ with $\zeta_A = r \xbar \zeta_A(x^B)$.  The radial component $\zeta_r$ is odd, while the angular components $\zeta_A$ are even.  The function $V$ is even. We repeat that $\xbar \lambda_A$ and  $\xbar \zeta_{A}$ will need further restrictions, discussed in Section \ref{sec:CanoniBoosts} below (see Eq. (\ref{eq:FormZeta0})).

\subsubsection{Symmetries}

The Poincar\'e vector fields read, in spherical coordinates,
\begin{align}
\xi & =br+a^{0},\\
\xi^{r} & =w_{1},\\
\xi^{A} & =Y^{A}+\frac{1}{r}\bar{D}^{A}w_{1}\,,
\end{align}
where $b \equiv b_i n^i$ and $w_1 \equiv  a^i \frac{\partial r}{\partial x^i}  = a^i n_i$ depend only on the angles and where $a^0$ is a constant.  The $Y^A$'s are the components of a Killing vector on the sphere with unit round metric $\xbar \gamma_{AB}$,
\be
 Y^A \frac{\partial}{\partial x^A} = \frac12 b_{ij} x^i \frac{\partial}{\partial x_j},  \qquad Y^A = \frac 12 b^{ij} Y^A_{ij}, 
\ee
where the $Y^A_{ij}=-Y^A_{ji}$'s form a basis of such Killing vectors and close in the Lie bracket according to the $so(3)$ algebra.  With the obvious redefinition $b_{ij} = \epsilon_{ijk} m^k$, one has
\be 
Y = m^1 Y_{(1)} + m^2 Y_{(2)} + m^3 Y_{(3)} \, , 
\ee
where
\begin{eqnarray}
Y_{(1)} &=& -\sin \varphi \frac{\partial}{\partial \theta} - \frac{ \cos \theta}{\sin \theta} \cos \varphi \frac{\partial}{\partial \varphi} \, , \\
Y_{(2)} &=& \cos \varphi \frac{\partial}{\partial \theta} - \frac{ \cos \theta}{\sin \theta} \sin \varphi \frac{\partial}{\partial \varphi} \, , \\
Y_{(3)} &=& \frac{\partial}{\partial \varphi} \, .
\end{eqnarray}

The function $b= b_i \frac{x^i}{r}$ is defined
on the $2$-sphere and reads
\be
b = b_1 \sin \theta \cos \varphi + b_2 \sin \theta \sin \varphi + b_3 \cos \theta,
\ee

The form 
$$\xi^{A}_{\textrm{translation}} = \frac{1}{r}\bar{D}^{A}w_{1}$$
 of the angular components $a^i \frac{\partial x^A}{\partial x^i}$ of the translations is most easily understood from the Killing equation $\mathcal{L}_\xi \gamma_{rA}= 0$ for the flat metric $\gamma_{ij}$ in spherical coordinates, which reads $(\partial_r \xi^B) r^2 \xbar \gamma_{AB} + \partial_A \xi^r = 0$, from which the result derives (there is no $O(r^0)$ piece in the translations).   One has explicitly for the three independent translations along  $x$, $y$, $z$ and the corresponding $w_1$'s, 
\begin{eqnarray}
\frac{\partial}{\partial x} &=&\sin \theta \cos \varphi \frac{\partial}{\partial r} + \frac1r \cos \theta \cos\varphi \frac{\partial}{\partial \theta} - \frac1r \frac{\sin \varphi}{\sin \theta} \frac{\partial}{\partial \varphi} \qquad (w_1 = \sin \theta \cos \varphi), \\
\frac{\partial}{\partial y} &=& \sin \theta \sin \varphi \frac{\partial}{\partial r} + \frac1r \cos \theta \sin \varphi \frac{\partial}{\partial \theta} + \frac1r \frac{\cos \varphi}{\sin \theta} \frac{\partial}{\partial \varphi} \qquad (w_1 = \sin \theta \sin \varphi), \\
\frac{\partial}{\partial z} &=& \cos \theta \frac{\partial}{\partial r} - \frac1r \sin \theta  \frac{\partial}{\partial \theta} \qquad (w_1 = \cos \theta ).
\end{eqnarray}
 
Useful equations fulfilled by $b$, $w_1$ and $Y^A$ are
\begin{gather}
\bar{D}_{A}\bar{D}_{B}b+\bar{\gamma}_{AB}b=0\,,\qquad\bar{D}_{A}\bar{D}_{B}w_{1}+\bar{\gamma}_{AB}w_{1}=0\,,\label{eq:Killing1}\\
\mathcal{L}_{Y}\bar{\gamma}_{AB}=Y^{C}\partial_{C}\bar{\gamma}_{AB}+\partial_{A}Y^{C}\bar{\gamma}_{BC}+\partial_{B}Y^{C}\bar{\gamma}_{AC}=0\,,\label{eq:Killing2}
\end{gather}

The gauge transformation parameters $\epsilon^\mu$ have the following asymptotic behaviour in spherical coordinates, $\epsilon = O(r^0)$, $\epsilon^{r}= O(r^0)$ and $\epsilon^{A} = O(r^{-1})$.

\section{Canonical realization of the boosts}
\label{sec:CanoniBoosts}

\subsection{Relativistic theories}

The fact that the Poincar\'e transformations leave the boundary conditions invariant simply means that these are well-defined within phase space and thus allowed transformations, mapping a phase point on a phase space point.  The validity of this  property is only the first step in the establishment of Poincar\'e invariance. 

 Relativistic invariance of the theory means indeed not only that phase space is mapped on itself by a Poincar\'e transformation, but also that the action is invariant.    We now turn to this second condition. 

As we have already emphasized many times, the Hamiltonian action is the sum of two terms, namely, the kinetic term which is linear in the time derivatives, and the Hamiltonian term, which does not depend on them. Invariance of the action under phase space transformations (not involving the time derivatives of the canonical variables) is equivalent  to separate invariance of the kinetic term and of the Hamiltonian.

Invariance of the kinetic term implies invariance of the underlying symplectic structure $\Omega = \int d^3 x d_V \pi^{ij} \wedge d_V h_{ij}$, where $d_V$ is the exterior derivative in field space.  This invariance expresses that the transformation is canonical and reads $\mathcal{L}_X \Omega = 0$, where $X$ is the vector field in field space corresponding to the transformation, and where $\mathcal{L}_X$ is the Lie derivative along $X$ in field space.  Using the formula $\mathcal{L}_X = \iota_X d_V + d_V \iota_X$ and $d_V \Omega = 0$, the invariance condition is easily seen to be equivalent to $d_V (\iota_X \Omega)= 0$ from which one derives $\iota_X \Omega = - d_V F$ for some generator $F$ determined up to an arbitrary constant.

The condition $d_V (\iota_X \Omega)= 0$ on the vector field $X$ is often called ``integrability condition for the generator $F$'' since the equation $d_V F = -\iota_X \Omega$ for $F$ is integrable if and only if the condition $d_V (\iota_X \Omega)= 0$ holds.

Once the existence of the canonical generator $F$ is established, the invariance of the Hamiltonian amounts to the condition $[F,H] = 0$. [If the transformation depends explicitly on time, the condition $[F,H] = 0$ becomes $\partial_t F + [F,H] =0$.  This is relevant to non-abelian symmetry algebras.]

The vector field $X$ defining the infinitesimal Poincar\'e transformation is given by (\ref{dh-Poincare}) and (\ref{eq:dp-Poincare}) with $\xi$ and $\xi^i$ given by (\ref{eq:KVMinkowski}).

\subsection{Integrability of the boost generators}

We want to check $d_V (\iota_X \Omega= 0)$ for the boosts, for which $a^0 = 0 = b_{ij} = a^k$ and so $\xi = b(x^A) r$, $\xi^k = 0$.   This is the most delicate point in showing that the theory is relativistic because
the boost vector fields grow linearly in $r$ and yield surface terms that are more intricate.  The other Poincar\'e transformations can be straightforwardly checked to be canonical. 

As we mentioned already, boosts, as all rigid symmetries, are defined up to a gauge symmetry. It turns out to be necessary for integrability to add to the boosts a gauge transformation with parameter
\be
\epsilon_{(b)} \equiv b F, \qquad (\Rightarrow \epsilon_{(b)} = 0 \textrm{ if } b = 0)  \label{eq:AccGauge}
\ee
where $F$ is a function of the fields to be given below. 
This is a transformation of order one. 
Being of order one, it is subleading with respect to $br$, but because it involves the fields, it has a non-trivial $d_V$ and  plays for that reason an essential role in ensuring integrability of the boost charges. This ``correcting'' gauge transformation is actually included here for that reason, to take care of a non-integrable term containing $d_V \xbar h_{AB}$ and $d_V \xbar h_{rr}$ (independent of $\lambda_A$) coming from the variation of the symplectic form under the original Poincar\'e boost, which it compensates.   
Other subleading terms fixed by $b$ will actually be needed below but it is not necessary to know them now because they do not affect integrability and contain only the momenta.  

With $(\xi, \xi^k)  = (b(x^A) r, 0)$, we find, upon partial integrations and using the fact that 
$$\int d^{3}x\,d_{V}\pi^{ij}\,d_{V}\pi_{ij}=\int d^{3}x\,d_{V}h^{ij}\,d_{V}h_{ij}=0, $$ that $d_{V}(\iota_{\xi}\Omega)$
reduces to a surface term,
\begin{eqnarray}
d_{V}(\iota_{\xi}\Omega) & = & \int\,d^{3}x\,[d_{V}(\delta_{\xi}\pi^{ij})\,d_{V}h_{ij}\,+\,d_{V}\pi^{ij}\,d_{V}(\delta_{\xi}h_{ij})]\,, \\
& = & \oint\,d^{2}S_{l}\,\Bigg[-\frac{1}{2}\xi\Big(d_{V}h^{ij}\nabla_{j}d_{V}h+d_{V}h\nabla_{j}d_{V}h^{ij}+d_{V}h_{jk}\nabla^{i}d_{V}h^{jk} \nonumber \\
 &  & \qquad\qquad\qquad\qquad-2d_{V}h_{jk}\nabla^{j}d_{V}h^{ik}+d_{V}h\nabla^{i}d_{V}h\Big)-\frac{1}{2}\nabla_{j}\xi d_{V}hd_{V}h^{ij}\Bigg]. \label{eq:IntegraA} 
  \end{eqnarray}

That there is no bulk contribution in $d_{V}(\iota_{\xi}\Omega)$ is actually a consistency condition that must hold.  It reflects the fact that the bulk Lagrangian is relativistic, so that the action is invariant under Poincar\'e transformations up to surface terms.  Failure to satisfy $d_{V}(\iota_{\xi}\Omega) = 0$  can therefore only arise from non-vanishing surface terms. The form of these surface terms depend evidently on the boundary conditions and therefore, these terms can be analysed only after precise boundary conditions have been given.

With strict parity conditions, the integrand of (\ref{eq:IntegraA}) would be even, yielding $d_{V}(\iota_{\xi}\Omega) = 0$ since $d^2 S_l \sim n_l$ is odd.  The non-vanishing of $d_{V}(\iota_{\xi}\Omega)$ arises from the fact that we adopt parity conditions involving a twist by improper gauge transformations.

Using the Killing equation for the boost parameter \eqref{eq:Killing1},
and the property that the Hamiltonian constraint is imposed up to the leading
order  $
\mathcal{G}^{(1)}\approx0
$,
we find that the surface integral reads
\begin{align}
d_{V}\left(\iota_{\xi}\Omega\right) & =-\int d\theta d\varphi\sqrt{\xbar{\gamma}}\Bigg[b\,d_{V}\xbar{h}d_{V}\Big(\xbar{h}_{rr}+\xbar{D}_{A}\xbar{\lambda}^{A}\Big)-\xbar{D}_{A}b\,d_{V}\xbar{\lambda}^{A}d_{V}\xbar{h}+b\xbar{D}^{A}d_{V}\xbar{\lambda}^{B}d_{V}\xbar{h}_{AB}\Big)\Bigg]\,,\label{eq:dW-P}
\end{align}
which is not zero.  Here, $\xbar h \equiv \xbar h_{AB} \xbar \gamma^{AB}$. The failure of $d_{V}(\iota_{\xi}\Omega)$ to vanish is a problem that must be cured since otherwise, the boosts would not be true symmetries of the theory.

The surface term at infinity (\ref{eq:dW-P}) is actually  exactly the same as in the full gravitational theory.    Thus although the Lorentz boosts do not coincide in the Pauli-Fierz theory and in the full Einstein theory, the former being the linearization of the latter, we fall back at this point on the already studied Einstein case and one can therefore repeat the arguments of \cite{Henneaux:2019yax}, to which we refer the reader.

There are two types of terms in $d_{V}\left(\iota_{\xi}\Omega\right)$: (i) terms that do not involve $\xbar \lambda_A$, namely $- b \, d_{V}\xbar{h}\, d_{V} \xbar{h}_{rr}$; (ii) terms linear in $\xbar \lambda_A$, which vanish when $\xbar \lambda_A= 0$.

We first consider the term independent of $\xbar \lambda_A$.  To compensate for it, we perform simultaneously with the Lorentz boost a gauge transformation with gauge parameter (\ref{eq:AccGauge}).  For such a $\epsilon_\mu$, one gets 
\begin{eqnarray}
d_{V}\left(\iota_{\epsilon}\Omega\right)&=&-\oint d^{2}S_{i}\left[d_{V}\epsilon\nabla_{j}\left(d_{V}h^{ij}-\gamma^{ij}d_{V}h\right)-\nabla_{j}d_{V}\text{\ensuremath{\epsilon}}\left(d_{V}h^{ij}-\gamma^{ij}d_{V}h\right)\right]\, , \\
&=& -2\int d\theta d\varphi\sqrt{\xbar{\gamma}}\, b\,d_{V}Fd_{V}\Big(\xbar{h}_{rr}+\xbar{D}_{A}\xbar{\lambda}^{A}\Big)\,.\label{eq:dW-G}
\end{eqnarray}
Adding both equations \eqref{eq:dW-P} and \eqref{eq:dW-G}, one has
\begin{eqnarray}
d_{V}\left(\iota_{\xi}\Omega\right)+d_{V}\left(\iota_{\epsilon}\Omega\right) &=& -\int d\theta d\varphi\sqrt{\bar{\gamma}}\Bigg[2b\,\Big(d_{V}F+\frac{1}{2}d_{V}\bar{h}\Big)d_{V}\Big(\bar{h}_{rr}+\bar{D}_{A}\bar{\lambda}^{A}\Big) \nonumber \\
&& \qquad \qquad -\bar{D}_{A}b\,d_{V}\bar{\lambda}^{A}d_{V}\bar{h}+b\bar{D}^{A}d_{V}\bar{\lambda}^{B}d_{V}\bar{h}_{AB}\Bigg]\,.
\end{eqnarray}
In order to get rid of the first term,  we set 
\be
F=-\frac{1}{2}\bar{h} \, ,
\ee
from which it follows that the right-hand side of  $d_{V}\left(\iota_{\xi}\Omega\right)+d_{V}\left(\iota_{\epsilon}\Omega\right)$ reduces to  
$$\int d\theta d\varphi\sqrt{\xbar{\gamma}}\Big[ \xbar{D}_{A}b\,d_{V}\xbar{\lambda}^{A}d_{V}\bar{h}-b\xbar{D}^{A}d_{V}\xbar{\lambda}^{B}d_{V}\xbar{h}_{AB}\Big]\, .$$
 This term vanishes if one sets $\xbar{\lambda}_{A}=0$.  Alternative ways to make $d_{V}\left(\iota_{\xi}\Omega\right)+d_{V}\left(\iota_{\epsilon}\Omega\right)$ vanish were explored in \cite{Henneaux:2019yax} and shown there to be equivalent to $\xbar \lambda_{A}=0$ up to proper (and thus admissible) gauge transformations\footnote{The situation is the following: by the gauge transformation with gauge parameters $\chi_r = 0$, $\chi_A = r \xbar \lambda_A$, one can set $\xbar \lambda_A = 0$.  But this gauge transformation has a generator that is on-shell equal to $\oint d \theta d \varphi \xbar \lambda_A \xbar \pi^{rA}$, which generically does not vanish in the absence of any condition on $\xbar \lambda_A$.  To achieve $\xbar \lambda_A = 0$ would appear therefore to need more than a mere proper gauge transformation in a theory where $\xbar \lambda_A$ would be unrestricted.  However,  the conditions on $\xbar \lambda_A$ studied in \cite{Henneaux:2019yax} that ensure integrability also make $\oint d \theta d \varphi \xbar \lambda_A \xbar \pi^{rA}$ vanish and hence enable one to set $\xbar \lambda_A$ equal to zero by a proper gauge transformation.}.
The condition $\xbar \lambda_A=0$ is the fourth ingredient in the asymptotic conditions announced at the opening of Section \ref{Asymptotia} above. 

To summarize, this additional requirement arises because the boosts are not automatically canonical transformations once we relax the conditions on the fields from strict parity conditions to twisted parity conditions.  The failure is a non-vanishing surface term. Something must be done in order to kill it and to ensure relativistic invariance.   This is why the condition:
\begin{enumerate}
\setcounter{enumi}{3}
\item The mixed radial-angular component of the metric vanishes to leading order
\be 
\xbar \lambda_A \equiv \xbar h_{rA} = 0
\ee
\end{enumerate}
is imposed from now on.

As direct consequences of this equation, we get from (\ref{eq:AsympLambda}) that $(\xbar \lambda_A)^{\textrm{odd}} = 0$ and that 
the form of the vector $\zeta_i$ in (\ref{TPh}) or in (\ref{eq:AsympLambda}) is restricted to fulfill
\be
\xbar \zeta_A = \xbar D_A \zeta_r \, , \label{eq:KeyRel0}
\ee
a condition that has important implications.

If we set $\zeta_r = U$ (with $U$ an odd function of the angles), we can rewrite (\ref{eq:KeyRel0}) as
\be
\zeta_r = \partial_r (rU), \qquad \zeta_A = \partial_A (rU) \label{eq:FormZeta0}
\ee
(recall that $\zeta_A = r \xbar \zeta_A$), or equivalently,
\be
\zeta_i = \partial_i(rU) \, . \label{eq:FormZeta}
\ee
It follows that (\ref{eq:AsympLambda}) is replaced by 
\be
\xbar{h}_{rr}=\text{even}\,,\quad \xbar{\lambda}_{A}=0 \,, 
\ee
and that (\ref{eq:AsympHAB}) becomes
\be
\xbar{h}_{AB}=(\xbar{h}_{AB})^{\text{even}}+2 (\xbar{D}_{A}\xbar D_{B} U +  \xbar{\gamma}_{AB}  U)\,, \qquad U(- \mathbf{n}) = -  U( \mathbf{n}) \, .
\ee

\subsection{Consequences}

\subsubsection{Expansion of time component of gauge transformations}
It is natural to split off the part of the gauge parameter $\epsilon$ that is dictated by the boost from the part that is free.  So we write
\be
\epsilon = \epsilon_{(b)} + T + O\left(\frac{1}{r}\right)\, , \qquad (\Leftrightarrow \xbar \epsilon = \epsilon_{(b)} + T )
\ee
where $T$ is a function of the angles only, independent from the boosts.  The function $T$ can have both even and odd parts under parity, but the odd part turns out to define a proper gauge transformation (zero surface integral, see below) so that only the even part is relevant.  The spherical harmonic decomposition of the relevant $T$ contains therefore only even harmonics and reads
\be
T = T_2 + T_4 + T_6 + \cdots
\ee
It starts at $T_2$ since the zero-mode  $T_0$,  which defines an isometry of Minkowski space, has no action on the spin-$2$ field and can therefore be factored out as explained above (time translations are included in $\xi$).  

\subsubsection{Boosts and spatial compensating gauge transformations}
The condition $\xbar \lambda_A = 0$ is generically not preserved under boosts when the action of the boosts on the canonical variables is defined as above.  One must add a compensating spatial gauge transformation $\epsilon^k_{(b)}$ that brings one back to $\xbar \lambda_A=0$.  This transformation  is easily worked out and again is the same as in the full Einstein theory. Its parameter $\epsilon^k_{(b)}$  takes the form
\be
\epsilon^r_{(b)} = 0\, , \qquad \epsilon^A_{(b)} = \frac{2b }{r \sqrt{\xbar{\gamma}}} \xbar \pi^{rA} \ ,
\ee
which must be added to the gauge transformation with parameter $
\epsilon_{(b)} = b F$ described above.  Adding this transformation does not spoil integrability of the boost charges, which simply receive a non-vanishing contribution from it.

It is again natural to split off the part of the gauge parameter $\epsilon^k$ that is dictated by the boost from the part that is free.  So we write
\be
\epsilon^k = \epsilon_{(b)}^k + W^k + O\left(\frac{1}{r}\right) \, , \qquad (\Leftrightarrow \xbar \epsilon^k = \epsilon_{(b)}^k + W^k )
\ee
where $W^k$ is a function of the angles only, independent from the boosts. 

The other Poincar\'e transformations preserve as such the condition $\xbar \lambda_A = 0$, without the need for compensating gauge terms.  This is obvious for the time translations as well as for the spatial rotations under which $\xbar \lambda_A $ transforms as a vector.  This is also true for the spatial translations, because these take the form  $ \xi^{r}  =w_{1}$, $\xi^{A}  =\frac{1}{r}\bar{D}^{A}w_{1}$ (with 
$w_1  = a^i n_i$).

\subsubsection{Geometrical implications of $\xbar \lambda_A=0$}

The condition $\xbar \lambda_A=0$ restricts the available gauge transformations $W^k$.  The same argument that led to (\ref{eq:FormZeta}) shows that the vector field $W^k$ must fulfill
\be
W_k = \partial_k (r W)\, , \label{eq:FormEpsilonK}
\ee
for some function of the angles $W(\mathbf n)$.  In polar coordinates, this yields
\be
W^r = W   , \qquad W^A = \frac{1}{r} \xbar D^A W \, .
\ee
As the function $T$, the function $W$ can have both even and odd parts under parity. But here, it is the even part that turns out to define a proper gauge transformation (zero surface integral, see below) so that only the odd part is relevant.  The spherical harmonic decomposition of the relevant $W$ contains therefore only odd harmonics and reads
\be
W = W_3 + W_5 + W_7 + \cdots
\ee
It starts at $W_3$ since the term $W_1$, which corresponds to background isometries (spatial translations) has no action on the spin-$2$ field when inserted in the gauge transformations and can be factored out.   

Vector fields of the form $\partial_k (r W)$  are quite special.  They are clearly hypersurface orthogonal.  The surfaces to which they are orthogonal are just given by $r W =$ constant.  This family of surfaces is invariant under dilations $r \rightarrow k r$, which map the surface $r W = C$ to the surface $rW = kC$.  Furthermore, if $W$ is odd, the surface $r W = C$ is mapped on the surface $rW = -C$ under the parity transformation $\mathbf{n} \rightarrow - \mathbf{n}$.
Finally, due to their specific $r$-dependence ($O(r^0)$ in cartesian coordinates), the commutator of two such vector fields is of lower order as $r \rightarrow \infty$.  The transformations ``asymptotically commute'' (more precisely, their commutator is an irrelevant proper gauge transformation).

Another interesting feature of the vector fields (\ref{eq:FormEpsilonK}) is that their angular components $\xbar \epsilon^A$ are completely determined by their radial component $\xbar \epsilon^r$.  In other words, a radial displacement $\xbar \epsilon^r$ is accompanied by a definite transformation of the two-sphere. Furthermore, this transformation has the very specific form of being a gradient, with $r =$ constant, $\xbar \epsilon^r$= constant being the lines to which the vector field $\xbar \epsilon^A$ is orthogonal\footnote{We stress that these transformations are subleading with respect to the homogeneous Lorentz transformations, for which there is no such connection.}.

The connection between radial and boundary displacements is to a limited extent reminiscent of the AdS/CFT correspondence where there is also an asymptotic relation between the two \cite{Henneaux:1985tv,Brown:1986nw}.  There, however, it is somewhat the opposite since it is the radial displacement that is determined by the boundary displacement, which must be a conformal Killing vector of the boundary.  Given the boundary displacement, the radial displacement is equal to the corresponding variation of the scale (the integrability conditions for the existence of the radial displacement lead, in fact, to the conformal Killing equations on the boundary).

\subsubsection{Transformation law of the leading orders of the fields}

It is useful, especially for computing the algebra of the charges below,  to collect at this point the variations of the leading orders of the fields under Poincar\'e transformations and gauge transformations, taking into account the above restrictions.  One finds
\begin{align}
\delta_{\xi, \epsilon} \xbar{h}_{rr} & =\frac{b}{\sqrt{\xbar{\gamma}}}\left(\xbar{\pi}^{rr}-\xbar{\pi}_{A}^{A}\right)+Y^{A}\partial_{A}\xbar{h}_{rr}\,,\label{eq:dhrr}\\
\delta_{\xi, \epsilon} \xbar{h}_{AB} & =\mathcal{L}_{Y}\xbar{h}_{AB}+2(\xbar{D}_{A}\xbar{D}_{B}W+\xbar{\gamma}_{AB}W)+\frac{b}{\sqrt{\xbar{\gamma}}}\left[2\xbar{\pi}_{AB}-\xbar{\gamma}_{AB}(\xbar{\pi}^{rr}+\xbar{\pi}_{C}^{C})\right] \nonumber \\
& \qquad \qquad + \frac{4}{\sqrt{\xbar{\gamma}}}\bar{D}_{(A}\left(b\xbar \pi_{B)}^{r}\right)\,,\\
\delta_{\xi, \epsilon} \xbar{\pi}^{rr} & =\mathcal{L}_{Y}\xbar{\pi}^{rr}+\sqrt{\xbar{\gamma}}\bigg[\frac{1}{2}b\left(6\xbar{h}_{rr}-\xbar{h}+\xbar{\triangle}\, \xbar{h}_{rr}\right)+\frac{1}{2}\xbar{D}_{A}b\left(2\xbar{D}_{B}\xbar{h}^{AB}-\xbar{D}^{A}\xbar{h}\right)  \nonumber \\
&\qquad \qquad - \frac12 \xbar{\triangle} (b \xbar h) +\xbar{\triangle} T\bigg]\,,\\
\delta_{\xi, \epsilon}\xbar{\pi}^{rA} & =\mathcal{L}_{Y}\xbar{\pi}^{rA}+\sqrt{\xbar{\gamma}}\bigg[\frac{1}{2}b\left(\xbar{D}_{B}\xbar{h}^{AB}-\xbar{D}^{A}\xbar{h}-2\bar{D}^{A}\xbar{h}_{rr}\right)+\frac{1}{2}\xbar{D}_{B}b\xbar{h}^{AB} \nonumber \\
& \qquad \qquad - \frac12\xbar{D}^{A} (b \xbar h) +\xbar{D}^{A} T \bigg]\,,\\
\delta_{\xi, \epsilon}\xbar{\pi}^{AB} & =\mathcal{L}_{Y}\xbar{\pi}^{AB}+\sqrt{\xbar{\gamma}}\Big\{\frac{1}{2}b\left(-\xbar{h}^{AB}+\xbar{\triangle}\, \xbar{h}^{AB}-2\xbar{D}^{(A}\xbar{D}_{C}\xbar{h}^{B)C}+\xbar{D}^{A}\xbar{D}^{B}\xbar{h}_{rr}+\xbar{D}^{A}\xbar{D}^{B}\xbar{h}\right) \nonumber \\
 & \qquad \qquad-\frac{1}{2}\xbar{D}_{C}b\left[-\xbar{D}^{C}\xbar{h}^{AB}+2\xbar{D}^{(A}\xbar{h}^{B)C}+\xbar{\gamma}^{AB}\left(\xbar{D}^{C}\bar{h}_{rr}+\xbar{D}^{C}\xbar{h}-2\xbar{D}_{F}\xbar{h}^{CF}\right)\right] \nonumber \\
 & \qquad  \qquad - \frac12 \xbar{\gamma}^{AB}\xbar{\triangle}(b\xbar h)  +\xbar{\gamma}^{AB}\xbar{\triangle}T+ \frac12 \xbar{D}^{A}\xbar{D}^{B}( b \xbar h) -\xbar{D}^{A}\xbar{D}^{B}T\Big\}\,,\label{eq:dpab}
\end{align}
with
\begin{align}
\mathcal{L}_{Y}\bar{h}_{AB} & =Y^{C}\partial_{C}\bar{h}_{AB}+\partial_{A}Y^{C}\bar{h}_{BC}+\partial_{B}Y^{C}\bar{h}_{AC}\,,\\
\mathcal{L}_{Y}\xbar{\pi}^{rr} & =\partial_{A}\left(Y^{A}\xbar{\pi}^{rr}\right)\,,\\
\mathcal{L}_{Y}\xbar{\pi}^{rA} & =-\partial_{B}Y^{A}\xbar{\pi}^{rB}+\partial_{B}\left(Y^{B}\xbar{\pi}^{rA}\right)\,,\\
\mathcal{L}_{Y}\xbar{\pi}^{AB} & =-\partial_{C}\xi^{A}\xbar{\pi}^{BC}-\partial_{C}\xi^{B}\xbar{\pi}^{AC}+\partial_{C}\left(\xi^{C}\xbar{\pi}^{AB}\right)\,.
\end{align}
We note that the zero mode of $T$ indeed drops from these formulas since $\xbar{\triangle}T_0 = 0 = \xbar{D}^{A}T_0$.  In the same way, the first spherical harmonic $W_1$ of $W$ would not contribute since $\xbar{D}_{A}\xbar{D}_{B}W_1+\xbar{\gamma}_{AB}W_1 = 0$.   As already observed in \cite{Geroch:1972up}, we also note that the spacetime translations do not affect the leading orders which transform non trivially only under homogeneous Poincar\'e transformations (boosts and rotations). 

Spacetime translations do modify the subleading orders, however. 
The transformation laws of the subleading orders are given for information  in Appendix
\ref{AppA}.

\subsection{More on the comparison with electromagnetism}

\subsubsection{Surface terms in the Pauli-Fierz action principle}

Unlike in electromagnetism, we do not
require extra surface fields to make the boosts integrable.  This is 
exactly as in the case of pure gravity.

A different light can be shed on this issue by examining the variational principle in hyperbolic coordinates, where time translations involve boosts.   In the case of electromagnetism, the variational principle  is ill-defined without the extra surface degrees of freedom in question \cite{Henneaux:2018gfi}.  This is an equivalent method for understanding the need for these additional fields.

It is thus natural to investigate the Pauli-Fierz variational principle in hyperbolic coordinates, where the Minkowski metric reads 
\be
	\label{eq:minkHyperbol}
	ds^2 = d\eta^2 + \eta^2 \tilde \gamma_{ab} dx^adx^b, 
\ee
where $\tilde \gamma_{ab} dx^adx^b$ denotes the metric on the unit hyperboloid $\mathcal{H}$,
\be
	\tilde \gamma_{ab} dx^adx^b = \frac{-1}{(1-s^2)^2}ds^2 + \frac{1}{1-s^2}  \xbar
	\gamma_{AB}dx^Adx^B.
\ee
We set $(x^\mu) =  (s, \eta, x^A)$ and $x^a=(s,x^A)$.   The covariant derivatives for tensor fields defined on $\mathcal{H}$ with metric $\tilde \gamma_{ab}$ are denoted by  $\mathcal D_a$.  

The variation of the Pauli-Fierz action written in general coordinates
\begin{equation}
S_{\text{PF}}=\int d^{4}x\sqrt{-g}\Big(-\frac{1}{4}\nabla_{\mu}h_{\nu\lambda}\nabla^{\mu}h^{\nu\lambda}+\frac{1}{2}\nabla_{\mu}h^{\mu\nu}\nabla_{\rho}h_{\nu}^{\rho}-\frac{1}{2}\nabla^{\mu}h\nabla^{\nu}h_{\mu\nu}+\frac{1}{4}\nabla_{\mu}h\nabla^{\mu}h\Big) \label{eq:PF-Action}
\end{equation}
reads
\begin{align}
\delta S_{\text{PF}} & =\frac{1}{2}\int d^{4}x\sqrt{-g}\left[\triangle h_{\mu\nu}+\nabla_{\mu}\nabla_{\nu}h-2\nabla_{\mu}\nabla_{\rho}h_{\nu}^{\rho}+g_{\mu\nu}\big(\nabla^{\lambda}\nabla^{\rho}h_{\lambda\rho}-\triangle h\big)\right]\delta h^{\mu\nu}\label{eq:LinEinstein}\\
 & \quad + B_{\text{PF}},\label{eq:BT}
\end{align}
where the boundary term is given by
\begin{equation}
B_{\text{PF}}=\int d^{4}x\sqrt{-g}\nabla_{\mu}\left[-\frac{1}{2}\nabla^{\mu}h_{\nu\lambda}\delta h^{\nu\lambda}+\big(\nabla_{\rho}h^{\rho\nu}-\frac{1}{2}\nabla^{\nu}h\big)\delta h_{\nu}^{\mu}-\frac{1}{2}\left(\nabla^{\nu}h_{\nu}^{\mu}-\nabla^{\mu}h\right)\delta h\right]\,.
\end{equation}
Setting $\delta S_{\text{PF}}=0$
yields the linearized Einstein equations in the bulk
\begin{equation}
\mathcal{E}_{\mu\nu}=\frac{1}{2}\left[\triangle h_{\mu\nu}+\nabla_{\mu}\nabla_{\nu}h-\nabla_{\mu}\nabla_{\rho}h_{\nu}^{\rho}-\nabla_{\nu}\nabla_{\rho}h_{\mu}^{\rho}+g_{\mu\nu}\left(\nabla^{\lambda}\nabla^{\rho}h_{\lambda\rho}-\triangle h\right)\right]=0\,,
\end{equation}
together with conditions at the boundary.

It is useful for later purposes to write the linearized Einstein
equations and the boundary term in terms of the (non gauge invariant) tensor $\mathcal{F}_{\,\,\,\mu\nu}^{\lambda}=\mathcal{F}_{\,\,\,\nu\mu}^{\lambda}$,
such that
\begin{eqnarray}
\mathcal{E}_{\mu\nu} & = & \nabla_{\lambda}\mathcal{F}_{\,\,\,\mu\nu}^{\lambda}\,,\\
B_{\text{PF}} & = & -\int d^{4}x\sqrt{-g}\nabla_{\lambda}\big(\mathcal{F}_{\,\,\,\mu\nu}^{\lambda}\delta h^{\mu\nu}\big)\,,\label{eq:BPF}
\end{eqnarray}
where
\begin{equation}
\mathcal{F}_{\,\,\,\mu\nu}^{\lambda}=\frac{1}{2}\Big[\nabla^{\lambda}h_{\mu\nu}-\Big(\nabla_{\rho}h_{(\mu}^{\rho}\delta_{\nu)}^{\lambda}-\frac{1}{2}\nabla_{(\mu}h\delta_{\nu)}^{\lambda}\Big)+g_{\mu\nu}\Big(\nabla_{\rho}h^{\lambda\rho}-\nabla^{\lambda}h\Big)\Big]\,.
\end{equation}
In terms of this tensor, the variation of the action principle simply reads
\begin{equation}
\delta S_{\text{PF}}=\int d^{4}x\sqrt{-g}\nabla_{\lambda}\mathcal{F}_{\,\,\,\mu\nu}^{\lambda}\delta h^{\mu\nu}-\int d^{4}x\sqrt{-g}\nabla_{\lambda}\big(\mathcal{F}_{\,\,\,\mu\nu}^{\lambda}\delta h^{\mu\nu}\big)\,.
\end{equation}

By writing the boundary term  (\ref{eq:BPF}) in hyperbolic coordinates $(\eta,s,x^{A})$, 
one gets
\begin{align}
B_{\text{PF}} & =B_{\text{PF}}^{\mathcal{H}}-\int d\eta\oint d^{2}x\sqrt{-g}\mathcal{F}_{\,\,\,\mu\nu}^{s}\delta h^{\mu\nu}\Big|_{s_{0}}^{s_{1}}\,,
\end{align}
with 
\begin{eqnarray}
&& B_{\text{PF}}^{\mathcal{H}}  =  - \lim_{\eta \rightarrow \infty} \intop_{\mathcal{H}}d^{3}x\sqrt{-g}\mathcal{F}_{\,\,\,\mu\nu}^{\eta}\delta h^{\mu\nu}\\
 && =  \lim_{\eta \rightarrow \infty} \intop_{\mathcal{H}}d^{3}x\sqrt{-\tilde{\gamma}}\eta^{3}\left[-\frac{1}{2}\nabla^{\eta}h_{\nu\lambda}\delta h^{\nu\lambda}+\big(\nabla_{\rho}h^{\rho\nu}-\frac{1}{2}\nabla^{\nu}h\big)\delta h_{\nu}^{\eta}-\frac{1}{2}\left(\nabla^{\nu}h_{\nu}^{\eta}-\nabla^{\eta}h\right)\delta h\right]\,, \qquad \label{eq:BPFH1}
\end{eqnarray}
where the integral is taken over the portion bounded by $s_0$ and $s_1$ of the 
the 3-dimensional unit hyperboloid $\mathcal{H}$.

The term at the time boundaries $s_0$ and $s_1$ will not be analyzed here, since its discussion depends on the chosen representation (whether one computes the amplitudes in the coordinate, the momentum, or any other representation), which we have not specified.  We shall focus only on the boundary term $B_{\text{PF}}^{\mathcal{H}}$ at spatial infinity.  

To analyse it, one needs boundary conditions at spatial infinity.  We adopt  (see
e.g., \cite{BeigSchmidt,Mann:2005yr,Compere:2011ve,Troessaert:2017jcm})
\begin{eqnarray}
h_{\eta\eta} & = & \frac{1}{\eta}\bar{h}_{\eta\eta}+\frac{1}{\eta^{2}}h_{\eta\eta}^{(2)}+o(\eta^{-2})\,,\\
h_{\eta a} & = & \bar{\lambda}_{a}+\frac{1}{\eta}h_{\eta a}^{(2)}+o(\eta^{-1})\,,\\
h_{ab} & = & \eta\bar{h}_{ab}+h_{ab}^{(2)}+o(1)\,.
\end{eqnarray}
This decay is compatible with the above decay as can be checked on the hyperplane $s=0$. 
The boundary term in (\ref{eq:BPFH1}) then becomes
\begin{eqnarray}
\hspace{-1cm} B_{\text{PF}}^{\mathcal{H}}&=&\frac{1}{2}\intop_{\mathcal{H}}d^{3}x\sqrt{-\tilde{\gamma}}\Big[\big(2\mathcal{D}_{b}\bar{h}_{a}^{b}-\mathcal{D}_{a}\bar{h}_{\eta\eta}-\mathcal{D}_{a}\bar{h}+8\bar{\lambda}_{a}\big)\delta\bar{\lambda}^{a}  \nonumber \\
&& \qquad  \quad +\big(2\mathcal{D}_{a}\bar{\lambda}^{a}+3\bar{h}_{\eta\eta}-\bar{h}\big)\delta\bar{h}_{\eta\eta}+\bar{h}^{ab}\delta\bar{h}_{ab}-\big(\mathcal{D}_{a}\bar{\lambda}^{a}+3\bar{h}_{\eta\eta}\big)\delta\bar{h}\Big]\,,  \label{eq:BT00}
\end{eqnarray}
where  $\bar{h}=\tilde{\gamma}^{ab}\bar{h}_{ab}$.  It is manifestly finite.

\subsubsection{The electromagnetic case}

The analogous computation in the case of electromagnetism, with the standard Maxwell action and  fall-off of the fields given in \cite{Campiglia:2017mua}
\begin{equation}
	\label{eq:EMasympHyperbol}
	A_a = \xbar A_a + \frac{1}{\eta} A_a^{(2)}  + o(\eta^{-1}),
	\qquad
	A_\eta = \frac{1}{\eta} \xbar A_\eta  + \frac{1}{\eta^2}A_\eta^{(2)} 
	+ o(\eta^{-2}),
\end{equation}
 leads to the boundary term 
\begin{equation}
B_{\text{em}}^{\mathcal{H}}= \int_{\mathcal H} d^3x \sqrt{-\tilde{\gamma}} \mathcal D^a \xbar A_\eta \delta
	\xbar A_a \, .
	\ee
If one were to impose that the standard Maxwell action was also stationary for arbitrary variations of $\xbar A_a$, one would get the equations $ \mathcal D^a \xbar A_\eta = 0$.  Among these equations, the equations $ \mathcal D^A \xbar A_\eta = 0$, obtained by varying $\xbar A_A$, force $\xbar A_\eta$ to be constant on the $2$-sphere.  This is not dictated by the bulk equations of motion, which allow non-trivial  $\xbar A_\eta$ (and also $\xbar A_A$) on the $2$-sphere.  It is thus desirable to have less stringent equations at the boundary. 

A way to realize this feature is given by the authors of \cite{Campiglia:2017mua},
who add the following boundary term to the action:
\begin{equation}
	\label{eq:EMactionHyperbol}
	S[A_\mu] = -\int d^4x \frac{\sqrt{-g}}{4} F_{\mu\nu} F^{\mu\nu} +
	\int_{\mathcal H} d^3x \sqrt{-\tilde{\gamma}} \, \xbar A_\eta (\mathcal D^a \xbar A_a
	+ \xbar A_\eta).
\end{equation}
Variation of the action now gives the boundary term
\be
{B'}_{\text{em}}^{\mathcal{H}} = \int_{\mathcal H} d^3x \sqrt{-\tilde{\gamma}} \Big(\mathcal D^a \xbar A_a + 2\xbar
	A_\eta\Big) \delta
	\xbar A_\eta
	\ee
on the hyperboloid.  Setting ${B'}_{\text{em}}^{\mathcal{H}}$ equal to zero yields the single equation 
\begin{equation}
	\label{eq:appextraEOM}
	\mathcal D^a \xbar A_a + 2 \xbar A_\eta = 0 \, ,
\end{equation}
which is a dynamical equation for the temporal component $ \xbar A_s$ (it takes the form $\partial_s \xbar A_s = \cdots$). 
This equation is in fact the leading term of the Lorenz gauge condition:
\begin{equation}
	\label{eq:appextraEOM2}
	\nabla^\mu A_\mu = \frac{1}{\eta^2} (\mathcal D^a \xbar A_a + 2 \xbar
	A_\eta) + o(\eta^{-2})
\end{equation}
and is thus compatible with Lorentz invariance.

This is the mechanism by which the leading term $\xbar A_s$ in the asymptotic expansion of the temporal component of the vector potential acquires dynamics and becomes a (surface) degree of freedom, e.g.  needing  initial conditions for its determination.  Once this new surface degree of freedom is introduced, there is a new global symmetry, which takes the form of an improper gauge transformation, and which combines with the improper gauge transformations of the spatial variables to yield the same angle-dependent $u(1)$ symmetries as the ones found at null infinity \cite{Henneaux:2018gfi}.

The Lorenz gauge condition is clearly compatible with all bulk equations of motion where it is just a gauge fixing condition.  Its asymptotic form   (\ref{eq:appextraEOM}) - which is the only thing being enforced anyway - freezes no improper gauge freedom and is thus acceptable.  That it freezes no improper gauge freedom follows from the fact that in order to reach it, one needs to perform a gauge transformation obeying a second order equation of motion, leaving arbitrary the asymptotic value of the gauge parameter $\xbar \varepsilon$ and its first order time derivative $\partial_s \xbar \varepsilon$ on the initial slice $s=0$ (say).  This is precisely the amount of improper gauge freedom (see \cite{Henneaux:2018gfi} for more information).  The asymptotic Lorenz gauge involves a definite choice of boundary Hamiltonian, relating how the improper gauge symmetry acts on different equal-$s$ slices.

\subsubsection{Back to the Pauli-Fierz case}

The electromagnetic formulation that we just recalled involves as asymptotic dynamical fields the coefficients $\xbar A_\eta$, $\xbar A_A$ of the leading terms of the spatial components $A_k$ of the vector potential (and their conjugates, not written here), which are also dynamical in the bulk, as well as the coefficient $\xbar A_s$ of the leading term of the temporal component $A_s$ of the vector potential, which is by contrast pure gauge in the bulk. 

Can one achieve a similar construction in the spin-$2$ case? To that end, in analogy with electromagnetism, one would like to add a boundary term to the action, in such a way that stationarity of the new action implies only four equations at the boundary, which should be dynamical equations for the leading terms $\xbar h_{ss}$, $ \xbar{\lambda}_s$ and $\xbar h_{sA}$ of the Lagrange multipliers (the analogs of $A_s$) and nothing else.  Again in analogy with electromagnetism, these equations should furthermore be the asymptotic part of the Poincar\'e invariant generalized De Donder gauge conditions.

The generalized De Donder gauge conditions take the form
\begin{equation}
G^{\mu}=\nabla_{\nu}h^{\mu\nu}+c\nabla^{\mu}h=0\,,\label{eq:CovGauge-1}
\end{equation}
with $c$ arbitrary.  Their fall-off in hyperbolic coordinates reads
\begin{eqnarray}
G^{\eta} & = & \frac{1}{\eta^{2}}\Big[\mathcal{D}_{a}\bar{\lambda}^{a}+(2-c)\bar{h}_{\eta\eta}-(1+c)\bar{h}\Big]+\mathcal{O}(\eta^{-3})\,,\label{eq:Getalambda1-1}\\
G^{a} & = & \frac{1}{\eta^{3}}\big(\mathcal{D}_{b}\bar{h}^{ab}+c\mathcal{D}^{a}\bar{h}_{\eta\eta}+c\mathcal{D}^{a}\bar{h}+3\bar{\lambda}^{a}\big)+\mathcal{O}(\eta^{-4})\,.\label{eq:Galambda1-1}
\end{eqnarray}
For the specific value $c=-1/2$, one recovers the de Donder gauge condition
\begin{equation}
G^{\mu}=\nabla_{\nu}h^{\mu\nu}-\frac{1}{2}\nabla^{\mu}h=0\,.\label{eq:CovGauge-2}
\end{equation}

Setting the boundary term (\ref{eq:BT00}) equal to zero (with unrestricted variations of the coefficients of the leading terms in the expansion of the fields)  leads to 10 equations, which is not the desired result.   In an attempt to cure this problem, one could try to repeat the  procedure that works for electromagnetism and add a surface term to the Pauli-Fierz action,
\begin{equation}
S_{\text{PF}} \rightarrow S=S_{\text{PF}}+B_{\text{PF}}\,,
\end{equation}
with
\begin{equation}
B_{\text{PF}}=\frac{1}{2}\intop_{\mathcal{H}}d^{3}x \sqrt{-\tilde \gamma}\Big[\bar{\lambda}_{a}\big(A\mathcal{D}_{b}\bar{h}^{ab}+B\mathcal{D}^{a}\bar{h}_{\eta\eta}+C\mathcal{D}^{a}\bar{h}\big)+F\bar{\lambda}_{a}\bar{\lambda}^{a}+G\bar{h}_{\eta\eta}^{2}+H\bar{h}_{\eta\eta}\bar{h}+J\bar{h}^{ab}\bar{h}_{ab}\Big]\,,\label{eq:GenB}
\end{equation}
in the hope that this would lead to satisfactory boundary equations.
The boundary term (\ref{eq:GenB}) is the most general one invariant under the homogeneous Lorentz transformations (the symmetries of the hyperboloid) and containing at most one derivative of the fields (to match the structure of the boundary term (\ref{eq:BT00}) and of the covariant gauges), It involves 7 arbitrary constants $\{A,B,C,F,G,H,J\}$.   This procedure does not work, however.

Indeed, the variation of the action acquires then the following form
\begin{align}
\delta S & =\int d^{4}x\sqrt{-g}\nabla_{\lambda}\mathcal{F}_{\,\,\,\mu\nu}^{\lambda}\delta h^{\mu\nu} \nonumber \\
 & \quad+\frac{1}{2}\intop_{\mathcal{H}}d^{3}x\sqrt{-\tilde{\gamma}}\Big\{\Big[(A+2)\mathcal{D}_{b}\bar{h}_{a}^{b}+(B-1)\mathcal{D}_{a}\bar{h}_{\eta\eta}+(C-1)\mathcal{D}_{a}\bar{h}+(2F+8)\bar{\lambda}_{a}\Big]\delta\bar{\lambda}^{a}\nonumber \\
 & \qquad\qquad\qquad\qquad\qquad+\Big[(-B+2)\mathcal{D}_{a}\bar{\lambda}^{a}+(2G+3)\bar{h}_{\eta\eta}+(H-1)\bar{h}\Big]\delta\bar{h}_{\eta\eta} \nonumber \\
 & \qquad\qquad\qquad\qquad\qquad+\Big[-A\mathcal{D}_{a}\bar{\lambda}_{b}-(C+1)\tilde{\gamma}_{ab}\mathcal{D}_{c}\bar{\lambda}^{c}+(H-3)\tilde{\gamma}_{ab}\bar{h}_{\eta\eta}+(2J+1)\bar{h}_{ab}\Big]\delta\bar{h}^{ab}\Big\} \nonumber \\
 & \quad-\Big[\int d\eta\oint d^{2}x\sqrt{-g}\mathcal{F}_{\,\,\,\mu\nu}^{s}\delta h^{\mu\nu}-\frac{1}{2}\oint d^{2}x\sqrt{-\tilde{\gamma}}\,\left(A\delta\bar{h}^{sa}\bar{\lambda}_{a}+B\delta\bar{h}_{\eta\eta}\bar{\lambda}^{s}+C\delta\bar{h}\bar{\lambda}^{s}\right)\Big]_{s_{0}}^{s_{1}}\,.
\end{align}
We can see that the action principle endowed with the boundary term
(\ref{eq:GenB}) leads to the following extra equations of motion
at the boundary $\mathcal{H}$
\begin{gather}
(-B+2)\mathcal{D}_{a}\bar{\lambda}^{a}+(2G+3)\bar{h}_{\eta\eta}+(H-1)\bar{h}=0\,,\label{eq:BEq1}\\
(A+2)\mathcal{D}_{b}\bar{h}_{a}^{b}+(B-1)\mathcal{D}_{a}\bar{h}_{\eta\eta}+(C-1)\mathcal{D}_{a}\bar{h}+(2F+8)\bar{\lambda}_{a}=0\,,\label{eq:BEq2}\\
-A\mathcal{D}_{(a}\bar{\lambda}_{b)}-(C+1)\tilde{\gamma}_{ab}\mathcal{D}_{c}\bar{\lambda}^{c}+(H-3)\tilde{\gamma}_{ab}\bar{h}_{\eta\eta}+(2J+1)\bar{h}_{ab}=0\,.\label{eq:BEq3}
\end{gather}
In order for the last equation to be empty, we request the coefficient of $\delta \xbar h_{ab}$ to be absent  (corresponding to the electromagnetic situation where $\delta \xbar A_a $ is absent in the final boundary term), which forces
$$
A = 0, \quad C= - 1, \quad H = 3, \quad J= - \frac12.
$$
But then the  equation (\ref{eq:BEq1}) becomes
$$(-B+2)\mathcal{D}_{a}\bar{\lambda}^{a}+(2G+3)\bar{h}_{\eta\eta}+2\bar{h}=0$$
while  (\ref{eq:BEq2}) reads
$$ 2\mathcal{D}_{b}\bar{h}_{a}^{b}+(B-1)\mathcal{D}_{a}\bar{h}_{\eta\eta}-2 \mathcal{D}_{a}\bar{h}+(2F+8)\bar{\lambda}_{a}=0
$$
and there is no way to choose the remaining coefficients so that these equations directly match (up to multiplicative constants) the covariant gauge conditions (\ref{eq:CovGauge-1}) (one gets $c = -1$ from the matching of the second equation with (\ref{eq:Galambda1-1}), but then $h$ drops from (\ref{eq:Getalambda1-1}) and matching of the first equation with  (\ref{eq:Getalambda1-1}) is impossible).  Accordingly, contrary to what can be done in electromagnetism, one cannot introduce independent boundary degrees of freedom ($\xbar h_{ss}$, $\xbar \lambda_s$ and $\xbar h_{sA}$) at infinity, obeying covariant dynamical equations, by adding an appropriate surface term to the action.  The procedure of enlarging the physical phase space along the lines of electromagnetism does not appear to be available.

As we have shown, however, this step is not necessary in the Pauli-Fierz case since a fully satisfactory formulation, exhibiting the full BMS symmetry, can be developed without surface degrees of freedom beyond the standard canonical variables. 

We close this section with two comments.  
\begin{itemize}
\item First, we note that the asymptotic equations (\ref{eq:BEq1})-(\ref{eq:BEq3}) are Poincar\'e invariant. Invariance under the homogeneous Lorentz transformations is indeed manifest in hyperbolic coordinates, while invariance under translations follows directly from the invariance of the leading terms in the asymptotic expansion (the translations affect only the subleading terms).  The problem is to match these equations with the asymptotic expansion of the generalized De Donder gauge conditions assumed to hold everywhere, in order to mimic the electromagnetic situation. Our negative result -- that this cannot be done -- raises interesting questions, e.g., can the matching be achieved up to a gauge transformation in the bulk?
\item Second, by allowing extra conditions on the asymptotic fields (which is a departure from the strict analogy with electromagnetism), one can implement the generalized De Donder gauge conditions in the bulk. For instance the condition $\xbar k = 0$ of \cite{Compere:2011ve}, which is equivalent to $\xbar h + 3 \xbar h_{\eta \eta} + 2 \mathcal{D}_a \lambda^a=0$, leads to equations that consistently incorporate the De Donder gauge when $B=-1$, $F=-1$ and $G=0$.  The extra condition $\xbar k = 0$ leads at the same time to the BMS$_4$ group as asymptotic symmetry group \cite{Compere:2011ve,Troessaert:2017jcm} (and not to any bigger one). Another possibility is to impose $ \lambda^a=0$ from the outset and take $H=3$ and $J = - 1/2$, which yields the generalized De Donder gauge conditions if $G = - \frac12 \left(\frac{c+7}{c+1}\right)$.  In other words, the De Donder gauge is of course perfectly consistent but there are subtleties in the variational principle.
\end{itemize}

\section{Charges within the linear theory}
\label{sec/CWithinL}

\subsection{Canonical generators of Poincar\'e transformations}
We treat separately the Poincar\'e symmetries and the (proper and improper) gauge symmetries.  We start with the Poincar\'e symmetries.

We follow in this section the standard procedure for deriving the Poincar\'e generators, where one supplements the relevant bulk integrals formed with the energy and momentum densities with the necessary boundary terms. We will derive in Section \ref{sec:WeakField_Charges} the same results by expanding the Poincar\'e generators of the full Einstein theory up to the pertinent weak field order.

The canonical generator of Poincar\'e transformations is given by
\begin{equation}
G_{\xi}=\int d^{3}x\left(\xi\mathcal{\mathcal{E}}+\xi^{i}\mathcal{P}_{i} + \epsilon_{(b)} \mathcal{G} +  \epsilon_{(b)}^i \mathcal{G}_i \right)+Q_{\xi}[\xi,\xi^{i}]\,,\label{eq:Gcan}
\end{equation}
where the surface term $Q_{\xi}$ has to be added in order that the
canonical generator $G_{\xi}$ fulfill
\be 
\iota_\xi \Omega = - d_V G_\xi  \label{eq:EqForGcan}
\ee
or, what is the same, in order that $G_{\xi}$
admits well-defined functional derivatives.

Since the boundary conditions make $\Omega$ finite and since Poincar\'e transformations preserve the boundary conditions, $\iota_\xi \Omega$ is finite and $G_\xi$ should also be finite.  However, the bulk term $\int d^{3}x\left(\xi\mathcal{\mathcal{E}}+\xi^{i}\mathcal{P}_{i}\right) $ in (\ref{eq:Gcan}) is
not obviously so and contains potential logarithmic divergences for
the spatial rotations $Y^{A}$ and the boosts $b$.  Our first task, then, is to verify explicitly that the volume integrals in $G_\xi$ are finite.   This is done in Appendix \ref{App:NoDiv}.

The ingredients that go into the proof are the same as the ones that guarantee finiteness of $\Omega$, and involve in particular that the leading orders of the constraints should vanish.  These conditions read, for the asymptotic fields in spherical coordinates, 
\begin{align}
\bar{D}_{A}\bar{\pi}^{rA}-\bar{\pi}_{A}^{A} &  = 0\,,\label{eq:Const1}\\
\bar{D}_{B}\bar{\pi}^{AB}+\bar{\pi}^{rA} &  = 0\,,\label{eq:Const2}\\
\bar{D}_{A}\bar{D}_{B}\bar{h}^{AB}-\bar{\triangle}\bar{h}_{A}^{A}-\bar{\triangle}\bar{h}_{rr} &  =0\,.\label{eq:Const3}
\end{align}

\subsubsection{Surface integrals}
Having checked that the bulk piece in $G_\xi$ is finite, we can now proceed to determine the accompanying surface integral.

The key equation for that matter is (\ref{eq:EqForGcan}), which can be rewritten as
\be
d_V Q_\xi = - \iota_\xi \Omega - d_V \Big[\int d^3x (\xi \mathcal{E} + \xi^i \mathcal{P}_i + \epsilon_{(b)} \mathcal{G} +  \epsilon_{(b)}^i \mathcal{G}_i ) \Big]
\ee
in view of (\ref{eq:Gcan}).

The computation is direct but cumbersome.  One integrates by parts the variation $d_V [\int d^3x (\xi \mathcal{E} + \xi^i \mathcal{P}_i + \epsilon_{(b)} \mathcal{G} +  \epsilon_{(b)}^i \mathcal{G}_i )] $ to bring it to the form $- \iota_\xi \Omega$.  In so doing one picks up surface terms, which are integrable as we have seen, and yield $d_V Q_\xi$ (some, but not all, of these terms are zero due to the generalized parity conditions). The procedure gives $Q_\xi$ up to an integration constant which we fix to vanish when all fields are zero.

We only write down here the final result, which is
\begin{equation}
Q_\xi=\int d\theta d\varphi\Bigg\{b\left[ \sqrt{\bar{\gamma}}(-\frac{1}{2}\bar{h}\bar{h}_{rr}+\frac{1}{4}\bar{h}^{2}-\frac{3}{4}\bar{h}_{AB}\bar{h}^{AB}) +\frac{2}{\sqrt{\bar{\gamma}}}\bar{\pi}_{A}^{r}\bar{\pi}^{rA} \right]+ 2Y_{A}\bar{\pi}^{rB}\bar{h}_{B}^{A}\Bigg\}\,.\label{eq:QXiFinal1}
\end{equation}
With the strict parity conditions of \cite{Regge:1974zd}, this surface term is zero and the charge reduces to the bulk integral\footnote{Correspondingly, the boost and rotation generators of the full theory, which are pure surface integrals,  contain no term quadratic in the fields and reduce to the linear terms  -- see \cite{Henneaux:2018cst,Henneaux:2018hdj,Henneaux:2019yax} and Section  \ref{sec:WeakField_Charges} below.}.  It is not so, however, as soon as one introduces a twist by an improper gauge transformation.  For the spacetime translations, described by Killing vectors that behave asymptotically with one power of $r$ less than the homogeneous Lorentz transformations,  there is no surface term and the generator reduces to the bulk piece only. 

\subsubsection{Poincar\'e charges} 

The Poincar\'e charges of the free massless spin-$2$ theory are obtained by adding the surface terms to the bulk contributions and are given by the following expressions:
\begin{itemize}
\item Energy-momentum
\be
a^0 E + a^i P_i = a^0 \int d^3 x \mathcal{E} + a^i \int d^3 x \mathcal{P}_i
\ee
\item Boost generators
\be
b_i M^{0i} = b_i \int d^3 x  x^i \mathcal{E} + b_i \oint_{S_2^{\infty}} d\theta d\varphi \frac{x^i}{r}\left[ \sqrt{\bar{\gamma}}(-\frac{1}{2}\bar{h}\bar{h}_{rr}+\frac{1}{4}\bar{h}^{2}-\frac{3}{4}\bar{h}_{AB}\bar{h}^{AB}) +\frac{2}{\sqrt{\bar{\gamma}}}\bar{\pi}_{A}^{r}\bar{\pi}^{rA} \right]
\ee
\item Angular momentum
\be
\frac12 b_{ij} M^{ij} = \frac12 b_{ij} \int d^3 x x^{[j} \mathcal{P}^{i]} + \frac12 b^{ij} \oint_{S_2^{\infty}} d\theta d\varphi  (2Y^{A}_{ij} \bar{\pi}^{rB}\bar{h}_{AB}) \, .
\ee
\end{itemize}

\subsection{Canonical generators of gauge transformations}

The canonical generators of the gauge transformations take the form

\be
G_{\epsilon, \epsilon^i}  =\int d^{3}x\left(\epsilon\mathcal{G}+\epsilon^{i}\mathcal{G}_{i}\right)+Q_{\epsilon,\epsilon^{i}}\,,\label{eq:Gg}\ee
where the surface integral is obtained from the equation $\iota_\epsilon \Omega = - d_V G_{\epsilon, \epsilon^i}$.  The bulk integral converges thanks to the faster fall-off condition on the constraints. As to the condition $\iota_\epsilon \Omega = - d_V G_{\epsilon, \epsilon^i}$, it reads in cartesian coordinates
\begin{equation}
d_V Q_{\epsilon,\epsilon^{i}}=\oint dS_{i}\left[\epsilon\nabla_{j}\left(d_V h^{ij}-\gamma^{ij}d_V h\right)-\nabla_{j}\text{\ensuremath{\epsilon}}\left(d_V h^{ij}-\gamma^{ij}d_V h\right)+2\epsilon_{j}d_V\pi^{ij}\right]\,.\label{eq:dQg}
\end{equation}
Note that the middle term is zero since $\nabla_i \epsilon \sim \frac{1}{r^2}$ and $h_{ij} \sim \frac{1}{r}$. In spherical coordinates, and using the asymptotic form of the momentum constraint, this expression becomes
\be
d_V Q_{\epsilon,\epsilon^{i}}  =   \oint d\theta d\varphi\Big[2\sqrt{\bar{\gamma}}T d_V \bar{h}_{rr}
+2W d_V\left(\bar{\pi}^{rr}-{\bar{\pi}^{A}}_A\right)\Big]\,,
\ee
with
\be
\epsilon = T + O\big(\frac{1}{r} \big), \qquad \epsilon^i = \partial^i(r W) + O\big(\frac{1}{r} \big) \,.
\ee
It is easily integrated to yield
\be
Q_{\epsilon,\epsilon^{i}}  =   \oint d\theta d\varphi\Big[2\sqrt{\bar{\gamma}}T  \bar{h}_{rr}
+2W \left(\bar{\pi}^{rr}-{\bar{\pi}^{A}}_A\right)\Big]\,,
\ee
where the integration constant has been adjusted so that the charges vanish for the zero field configuration.  Putting all terms together, we thus get
\be
G_{\epsilon,\epsilon^i}  =\int d^{3}x\big(\epsilon\mathcal{G}+\epsilon^i \mathcal{G}_{i} \big)+\oint d\theta d\varphi\Big[2\sqrt{\bar{\gamma}}T  \bar{h}_{rr}
+2W \left(\bar{\pi}^{rr}-{\bar{\pi}^{A}}_A\right)\Big]\,,\label{eq:Gg2}\ee

Two comments are in order:
\begin{itemize}
\item Since $\bar{h}_{rr}$ is strictly even and $\bar{\pi}^{rr}-{\bar{\pi}^{A}}_A$ strictly odd, only the event part of $T$ and the odd part of $W$ contribute to the charges.  The parts with opposite parity define proper gauge transformations, as  announced.
\item The zero mode $T_0$ of $T$, if included,  would drop out from the generator $G_{\epsilon,\epsilon^i} $ since in the linearized theory one has  $\int d^{3}x T_0 \mathcal{G} + \oint d\theta d\varphi [2\sqrt{\bar{\gamma}}T_0  \bar{h}_{rr}] \equiv 0$ (identically and not just on-shell), as can be seen by converting the surface integral into a volume integral using Stokes' theorem.   This is of course as it should since $T = T_0$ has no action on the fields. A similar remark holds for the first spherical harmonic $W_1$ of  $W$.  We shall come back to this point in the next section.
\end{itemize}

\section{Structure of the algebra of the charges}
\label{sec:Algebra}

\subsection{Poisson bracket algebra}

The Poincar\'e charges close in the Poisson bracket according to the Poincar\'e algebra.  The improper gauge transformations (pure supertranslations) form an infinite-dimensional abelian algebra.  Finally, the pure supertranslations commute with the translations and form a representation of the homogeneous Lorentz group given by
\begin{equation}
	\Big[ b_i M^{0i} + \frac12 b_{ij} M^{ij} , G_{\epsilon^\perp, \epsilon^i} \Big] =
	G_{\hat{\epsilon}^\perp, \hat{\epsilon}^i}\, ,
\end{equation}
where $\hat{\epsilon}^\perp$ and  $\hat{\epsilon}^i$ are asymptotically given by
\be
\hat{\epsilon}^\perp = \hat{T} + O\left(\frac{1}{r}\right) , \qquad \hat{\epsilon}^i = \partial^i (r \hat{W}) + O\left(\frac{1}{r}\right) \, ,
\ee
with
\be
		\hat T  = Y^A\partial_A T - 3 b W - \partial_A b \xbar D^A W - b
	\xbar D_A\xbar D^A W , \qquad
	\hat W  = Y^A \partial_A W - b T . \label{eq:LorentzTransTW}
\ee
The continuation of this asymptotic behaviour as one marches inside depends on the choice of $\epsilon^\perp$ and  $\epsilon^i$ in the bulk but is in any case irrelevant since one can always add proper gauge transformations to any transformation without changing its physical content.

The formula (\ref{eq:LorentzTransTW}) gives in general a non-vanishing first spherical harmonic contribution to $W$ even if $T = T_2 + T_4 + \cdots$ and $W = W_3 + W_5 + \cdots$.  This term can be subtracted by hand in (\ref{eq:LorentzTransTW}) or can be kept since the equivalence $W \sim W + u_1$ is automatically implemented in the formula for the generators of the improper gauge symmetries of the linear theory.

As shown in \cite{Troessaert:2017jcm}, one can make a change of basis $(T, W) \rightarrow (\tau)$ in the space of the pure supertranslations in order to bring (\ref{eq:LorentzTransTW}) to the more familiar form
\be
\hat{\tau} = Y^A \partial_A \tau - \partial^Ab\partial_A \tau -  b \tau  \label{eq:LorentzTransTW2}
\ee
(see also \cite{Henneaux:2018cst}). 
Our purpose now is to shed light on the pure supertranslation representation (\ref{eq:LorentzTransTW2})  of the homogeneous Lorentz group and to investigate how it connects with the four-dimensional translation representation.

\subsection{Homogeneous Lorentz group and the two-sphere at spatial infinity}
At large spatial distances, the Poincar\'e transformations are dominated by their terms linear in $r$. Spacetime translations are therefore subdominant, and furthermore, in the boosts, the relevant term is $rb$. 

The  asymptotic parametrization of the homogeneous Lorentz group is  given by the rotation Killing vectors $Y^A $  
\be
Y^A \frac{\partial}{\partial x^A} = \frac12 b_{ij} x^i \frac{\partial}{\partial x_j}, 
\ee
and the function $b$  
\be
 b= b_i \frac{x^i}{r} .
\ee

Now, one can form from $b$ a vector field $B^A$ tangent to the $2$-sphere,
\be
 B^A = - \xbar \gamma^{AB} \partial_B b = \sum_{i=1}^3 b_i B_{(i)}^A
\ee
with
\begin{eqnarray}
B_{(1)} &=&  -\cos \theta \cos \varphi \frac{\partial}{\partial \theta} + \frac{ \sin \varphi}{\sin \theta}  \frac{\partial}{\partial \varphi} \, , \\
B_{(2)} &=& -\cos \theta \sin \varphi \frac{\partial}{\partial \theta} - \frac{ \cos \varphi}{\sin \theta}  \frac{\partial}{\partial \varphi} \, ,\\
B_{(3)} &=&  \sin \theta\frac{\partial}{\partial \theta}\, .
\end{eqnarray}
The equation (\ref{eq:Killing1}) fulfilled by $b$, namely, 
$\bar{D}_{A}\bar{D}_{B}b+\bar{\gamma}_{AB}b=0$, 
shows that the $B^A_{(i)}$'s are conformal Killing vectors,
\be
\xbar D_A B_B + \xbar D_B B_A =  2 b \xbar \gamma_{AB} \, .
\ee
Furthermore, the Lie bracket algebra of the rotation vector fields $Y^A$ and the conformal vector fields $B^A$ is precisely the Lorentz algebra (in our sign conventions),
\be
[Y_{(i)}, Y_{(j)}] =  - \epsilon_{ijk} Y_{(k)}, \qquad [Y_{(i)}, B_{(j)}] = -\epsilon_{ijk} B_{(k)}, \qquad [B_{(i)}, B_{(j)}] = \epsilon_{ijk} Y_{(k)}.
\ee
This is of course the familiar realization of the Lorentz algebra as conformal algebra of the round $2$-sphere.  We stress, however, that here the $2$ sphere in question is the $2$-sphere at spatial infinity and not the celestial sphere at null infinity.

We also note that the function $b'$ from which the Lie bracket $[Y, B] =B'$ derives  is simply given by
\be
B_A = - \partial_A b \qquad \Rightarrow \qquad [Y,B]_A = - \partial_A b', \quad b' = Y^A \partial_A b
\ee

\subsection{Representations}

A well-known infinite family of representations of the Lorentz algebra defined in the vector space of functions on the $2$-sphere is of direct relevance to our analysis.

Let $\tau (\theta, \varphi)$ be a function on the $2$-sphere.  We define
\be
\delta_{Y, b} \tau  \equiv \delta_{Y + B } \tau = \mathcal{L}_{Y+B} \tau - k b \tau = (Y^A + B^A)\partial_A \tau - k b \tau \, . \label{eq:RepS0}
\ee
For any real number $k$, (\ref{eq:RepS0}) defines a representation of the homogeneous Lorentz algebra, which we denote $\Lambda_k$.  One has indeed
$$
[\delta_{Y + B}, \delta_{Y'+ B'}] \tau= \delta_{[Y+B, Y'+B']} \tau 
$$
for any $k$ (we restrict $k$ to be real so that $\delta_{Y, b} \tau$ is real when $\tau$ is real).
 This is the infinitesimal version of the representation of the homogeneous Lorentz group given by
$$
\tau(x) \rightarrow \tau'(x') = \tau(x) \Delta(x)^k \, ,
$$
where $k$ is the ``conformal weight'' of $\tau$  and $\Delta(x)^2$ the conformal factor induced by the conformal transformation $x^A \rightarrow {x'}^A$.

These representations are not irreducible when $k$ is a non-negative integer.  The subspace $P_k$ of dimension $(k+1)^2$ spanned by the spherical harmonics $Y_{lm}$ with $l \leq k$ is indeed invariant.  That this is so is clear under rotations since the spherical harmonics for fixed $l$ form representations of the rotation group.   Invariance of $P_k$ under boosts  can then most easily be seen by recalling that $P_k$ can be viewed as the space of polynomials of degree $\leq k$ in the components $n^i$ of the unit normal to the sphere (subject to the relation $n^i n_i = 1$).   With $b= b_i n^i = b^i \frac{x^i}{r}$ and $\tau = t_{i_1 i_2 \cdots i_p} n^{i_1} n^{i_2} \cdots n^{i_p} = t_{i_1 i_2 \cdots i_p} \frac{x^{i_1} x^{i_2} \cdots x^{i_p}}{r^p}$ ($t_{i_1 i_2 \cdots i_p}= t_{(i_1 i_2 \cdots i_p)}$), one gets
\begin{eqnarray}
\delta_b \tau &=& - \xbar{\gamma}^{AB} \partial_A b \partial_B \tau - k b \tau  \nonumber \\
&=& - \xbar \gamma^{AB} {e_A}^i \partial_i b \, {e_B}^j \partial_j \tau - k b \tau \nonumber \\
&=& - p \xbar \gamma^{AB} {e_A}^i \frac{b_i}{r}\, {e_B}^j \frac{t_{jj_2 \cdots j_p}x^{j_2} \cdots x^{j_p}}{r^p} - k b \tau  \nonumber \\
&=& - \frac{p}{r^{p-1}} \Big(\frac{\xbar \gamma^{AB}}{r^2} {e_A}^i {e_B}^j \Big)b_i \,  t_{jj_2 \cdots j_p}x^{j_2} \cdots x^{j_p} - k b \tau \nonumber \\
&=& - \frac{p}{r^{p-1}} \Big( \gamma^{AB} {e_A}^i {e_B}^j + n^i n^j \Big)b_i \,  t_{jj_2 \cdots j_p}x^{j_2} \cdots x^{j_p} + p b \tau - k b \tau \nonumber \\
&=& - p \, b^j  t_{jj_2 \cdots j_p}n^{j_2} \cdots n^{j_p} + (p-k) b \tau \, . \label{eq:S1}
\end{eqnarray}
Here,  the ${e_A}^i$'s are the tangent vectors to the $2$-sphere,
$$
{e_A}^i = \frac{\partial x^i}{\partial x^A},
$$
 and we have used the relationship $\gamma^{AB} {e_A}^i {e_B}^j + n^i n^j = \delta^{ij}$.  If $\tau$ is a polynomial of degree $p$, the first term in (\ref{eq:S1}) is a polynomial of degree $p-1$ while the product $b \tau$ in the second term is a sum of a polynomial of degree $p-1$ and a polynomial of degree $p+1$ (the polynomial of degree $p-1$ arises from the term proportional to the metric in $b \tau$  and the equation $n^i n_i = 1$). We thus see that $\delta_b \tau$ itself is the sum of a polynomial of degree $p-1$ and a polynomial of degree $p+1$.  This latter polynomial is absent if $p=k$, which proves the assertion that the degree remains $\leq k$ when $p \leq k$.
 
 The representation in the invariant subspace $P_k$ is of course just equivalent to the representation $S_k$ of the homogeneous Lorentz group by symmetric, traceless tensors $t_{\alpha_1 \cdots \alpha_k}$, containing once and only once all the  representations of the rotation group with integer spins $\leq k$.  The representation $\Lambda_k / S_k$  in the quotient space is infinite-dimensional and contains higher spins.
 It should be stressed that although the representation space of $\Lambda_k$ decomposes as a direct sum of $P_k$ and of the vector space containing the higher spherical harmonics, the representation $\Lambda_k$ itself is not the direct sum of $\Lambda_k /S_k$ and $S_k$ since the transformation of a spherical harmonic of degree $k+1$ under boosts involves generically a spherical harmonic of degree $k$ (see the analysis for $k=1$ in \cite{Sachs:1962zza}).
 
\subsection{Translations and pure supertranslations}
The $k=1$ representation is of particular interest since the four-dimensional invariant subspace $P_1$ is  the space of the translations, which transform in the four-dimensional representation $S_1$ of the homogeneous Lorentz group.  A generic translation is given by
\be
a^0 \frac{\partial }{\partial t} + a^i \frac{\partial }{\partial x^i}\, .
\ee
The translations in time correspond to the spin-$0$ representation of the rotation group, while the space translations correspond to the spin-$1$ representation.  

If we decompose a polynomial $\tau$ of degree $\leq 1$ ($\in P_1$) into even and odd parts,
\be
\tau = T_0 + W_1\, , \qquad T_0 = a^0\, , \qquad W_1=  a_i n^i\, ,
\ee
one can readily check that the transformation rule $\delta \tau =  Y^A \partial_A \tau - \partial^A b\partial_A \tau -  b \tau$ reproduces the standard transformation law of the translation generators under spatial rotations and boosts.

We have also seen that the pure supertranslations transform in the representation $\Lambda_1 /S_1$.  We have thus all the ingredients to form the full infinite-dimensional representation $\Lambda_1$.  And indeed, when the Einstein interaction is switched on, the translations ``fill the holes'' left empty by the ``zero modes'' $(T_0, W_1)$ of the pure supertranslations, which are pure gauge in the linear theory.  These become non-trivial and equal to the energy and momentum in the full theory.   This mechanism is algebraically possible because the translations transform in the right way, and because the pure supertranslations also transform under boosts as they should, inducing the correct ``zero mode'' term. This term is equal to a surface term involving the next order of the metric in the weak field expansion, which takes the precise form that allows exact matching. 

Note that the generators of the pure supertranslations get corrected in the process: the constraints of the linear theory are replaced by the constraints of the full theory, and the surface terms at infinity involve the complete fields.

\subsection{Electromagnetism and angle-dependent $u(1)$ transformations}

The electromagnetic situation shares some similarities with the gravitational one. 

The improper gauge symmetries of free electromagnetism transform in the infinite-dimensional representation $k=0$.  More precisely, they transform in $\Lambda_0 / S_0$ since the zero mode of the angle-dependent $u(1)$ improper gauge transformations (which defines the invariant subspace of the one-dimensional trivial representation $S_0$),  actually vanishes in the absence of charges.  

Coupling to charged matter ``fills the hole'' and leads to the full representation $\Lambda_0$.
This is a non-trivial step, however.  It has recently been shown that serious difficulties exist for this mechanism to apply to Yang-Mills type couplings among free abelian gauge fields, which appear to eliminate the infinite-dimensional symmetry at spatial infinity \cite{Tanzi:2020fmt}.  So while couplings to charged matter  fields leads to the representation $\Lambda_0$, exhibiting an algebraic situation similar to the gravitational one, self-couplings would destroy the symmetry and appear thus to be quite distinct.

\section{Charges from the weak field expansion of gravity}
 \label{sec:WeakField_Charges}

We have derived above the charges of the free Pauli-Fierz theory without referring to the full Einstein theory.  This derivation is useful in order to get insight into the structure of the charges of theories of higher spin gauge fields in Minkowski space, for which the analog of the full Einstein theory does not exist in closed form. 

In the spin-$2$ case considered here, however, one can proceed more effectively by starting from the charges of the full Einstein theory and linearizing them.  The procedure is also instructive as it shed light on the connection between the Poincar\'e and gauge charges.  This task is carried out in this section.   

The key equations are the weak field expansions of the canonical variables\footnote{There are two important expansions here: one is the weak field expansion controlled by $\kappa$, which can be related to the gravitational coupling constant $G$; the other is the asymptotic expansion at infinity in powers of $\frac{1}{r}$.  We  use in this section the following convention: the total fields are decorated with the subscript ``Full''. Each field appearing in the weak field expansion has furthermore an asymptotic expansion, e.g., $ h_{ij} = \frac{\xbar h_{ij}}{r} + O(\frac{1}{r^2})$, $ \varphi_{ij} = \frac{\xbar \varphi_{ij}}{r} + O(\frac{1}{r^2})$.}
\be
g_{ij} = \delta_{ij} + \kappa h_{ij} + \kappa^2  \varphi_{ij} +O(\kappa^3), \qquad \pi^{ij}_{\textrm{Full}} = \kappa \pi^{ij} + \kappa^2 p^{ij} + O(\kappa^3) ,  \label{eq:WeakF2}
\ee
and of the asymptotic symmetries of the full theory
\begin{eqnarray}
&& \xi_{\textrm{Full}}^\perp = \xi_{\textrm{Poincar\'e}}^\perp +  \kappa \epsilon^\perp   + O(\kappa^2),  \qquad \epsilon^\perp = T + O\Big(\frac{1}{r} \Big) \\
&&  \xi^i_{\textrm{Full}} = \xi^i_{\textrm{Poincar\'e}} + \kappa \epsilon^i  + O(\kappa^2), \qquad \epsilon^i = \partial^i (Wr) + O\Big(\frac{1}{r} \Big).
\end{eqnarray}
as well as the expansion of the constraints
\be
\mathcal{G}_{\textrm{Full}} = \kappa \mathcal{G}[h_{ij}] + \kappa^2 \Big(\mathcal{E}[h_{ij}, \pi^{ij} ] + \mathcal{G}[\varphi_{ij}] \Big) + O(\kappa^3)
\ee
and
\be
\mathcal{G}_{k, \textrm{Full}} = \kappa \mathcal{G}_k[\pi^{ij}] + \kappa^2 \Big(\mathcal{P}_k[h_{ij}, \pi^{ij}] + \mathcal{G}_k[p^{ij}] \Big) + O(\kappa^3)
\ee
From $\mathcal{G}_{k, \textrm{Full}} \approx 0$ and $\mathcal{G}_{k, \textrm{Full}} \approx 0$ we get the  equations
\begin{eqnarray}
&& \mathcal{G}[h_{ij}] \approx 0, \qquad \mathcal{E}[h_{ij}, \pi^{ij}] + \mathcal{G}[\varphi_{ij}]  \approx 0 , \qquad \textrm{etc} \\
&& \mathcal{G}_k[\pi^{ij}] \approx 0 , \qquad \mathcal{P}_k[h_{ij}, \pi^{ij}] + \mathcal{G}_k[p^{ij}] \approx 0, \qquad \textrm{etc} 
\end{eqnarray}
It will be convenient to denote the deviation from the flat metric by $h_{ij}^{\textrm{Full}}$, i.e., 
\be
h_{ij}^{\textrm{Full}} \equiv g_{ij} - \delta_{ij} = \kappa h^{ij} + \kappa^2  \varphi^{ij} +O(\kappa^3) \ .
\ee

Consider first the BMS supertranslation generator $G^{\textrm{SupTrans}}_{\textrm{Full}}$ in the full theory.  These are given, for the parity conditions with an improper twist,  by \cite{Henneaux:2018hdj,Henneaux:2019yax}
\be
G^{\textrm{SupTrans}}_{\textrm{Full}} = \int d^3x \Big(\xi_{\textrm{Full}}^\perp \mathcal{G}_{\textrm{Full}} + \xi^k_{\textrm{Full}} \mathcal{G}_{k, \textrm{Full}} \Big) + \oint d\theta d\varphi\Big[2\sqrt{\bar{\gamma}}T  \xbar{h}_{rr}^{\textrm{Full}}
+2W \left(\bar{\pi}^{rr}-{\bar{\pi}^{A}}_A\right)_{\textrm{Full}}\Big]\
\ee
where $(\xi_{\textrm{Full}}^\perp, \xi^k_{\textrm{Full}} )$ do not include Lorentz boosts or spatial rotations and so are $O(r^0)$, i.e. remain bounded at infinity.  Furthermore, in this formula, $T$ includes a zero mode $T_0 \equiv a^0$ and $W$ includes a first spherical harmonic $W_1 \equiv w_1$ with $w_1 \equiv a^k n_k$ (translations are included). Plugging the weak field expansion into this expression, with $T \rightarrow a^0 + \kappa (T- T_0)$ to conform with the weak field expansion of the symmetries ($T-T_0$ is arbitrary and can thus be rescaled), as well as $W \rightarrow w_1 + \kappa (W-W_1)$,  one gets a priori both terms of order one:
\be
G^{\textrm{SupTrans}}_{(1)} = \int d^3x \Big(a^0 \mathcal{G}[h_{ij}]  + a^k \mathcal{G}_k[\pi^{ij}]  \Big) + \oint d\theta d\varphi\Big[2\sqrt{\bar{\gamma}}a^0  \xbar{h}_{rr}
+2w_1 \left(\bar{\pi}^{rr}-{\bar{\pi}^{A}}_A\right)\Big]
\ee
 and of order two, 
\begin{eqnarray}
G^{\textrm{SupTrans}}_{(2)} &=& \int d^3x \Bigg\{a^0 \Big(\mathcal{E}[h_{ij},\pi^{ij}] + \mathcal{G}[\varphi_{ij}] \Big)  + a^k \Big(\mathcal{P}_k[h_{ij}, \pi^{ij}] + \mathcal{G}_k[p^{ij}] \Big)  \Bigg\}\nonumber \\
&& + \int d^3x \Bigg\{ (T-T_0) \mathcal{G}[h_{ij}]  + \partial^k(W-W_1) \mathcal{G}_k[\pi^{ij}]\Bigg\} 
\nonumber \\
&& + \oint d\theta d\varphi\Big[2\sqrt{\bar{\gamma}}a^0  \xbar{\varphi}_{rr} + 2\sqrt{\bar{\gamma}}(T- T_0)  \xbar{h}_{rr}\Big] \nonumber \\
&& +\oint d\theta d\varphi\Big[ 2w_1 \left(\bar{p}^{rr}-{\bar{p}^{A}}_A\right) + 2(W-W_1) \left(\bar{\pi}^{rr}-{\bar{\pi}^{A}}_A\right)\Big]\, .
\end{eqnarray}

Using Stoke's theorem, the term of order one is easily seen to identically vanish because of the identities (written in cartesian coordinates)
\begin{eqnarray}
&& \mathcal{G}[h_{ij}] + \partial^{i} \partial^j \big(\delta_{ij} h - h_{ij}\big)=0 \quad \Rightarrow \quad \int d^3x a^0 \mathcal{G}[h_{ij}] + \oint d^2S_i  a^0\partial_j \big(\delta^{ij} h - h^{ij}\big) =0 \\
&& \mathcal{G}_k[\pi^{ij}] + 2 \partial_i {\pi^{i}}_{k} = 0  \quad \Rightarrow \quad \int d^3x a^k \mathcal{G}_k[\pi^{ij}] + 2 \oint d^2 S_i a^k {\pi^{i}}_{k} = 0
\end{eqnarray}
and $ \oint d^2S_i  \partial_j \big(\delta^{ij} h - h^{ij}\big) = 2 \oint d\theta d\varphi\sqrt{\bar{\gamma}}  \xbar{h}_{rr} $, $a^k \oint d^2 S_i  {\pi^{i}}_{k} = \oint d\theta d\varphi w_1 \left(\bar{\pi}^{rr}-{\bar{\pi}^{A}}_A\right)$, 
 while the term of order two can be transformed, using the same identities but now for $\varphi_{ij}$ and $p^{ij}$, into 
\begin{eqnarray}
G^{\textrm{SupTrans}}_{(2)} &=& \int d^3x \Bigg\{a^0 \mathcal{E}[h_{ij},\pi^{ij}]   + a^k \mathcal{P}_k[h_{ij}, \pi^{ij}]   \Bigg\} \nonumber  \label{eq:TransFromFull} \\
&& + \int d^3x \Bigg\{ (T-T_0) \mathcal{G}[h_{ij}]  + \partial^k(W-W_1) \mathcal{G}_k[\pi^{ij}]\Bigg\} 
\nonumber \\
&& + \int d\theta d\varphi\Big[ 2\sqrt{\bar{\gamma}}(T- T_0)  \xbar{h}_{rr}  + 2(W-W_1) \left(\bar{\pi}^{rr}-{\bar{\pi}^{A}}_A\right)\Big] \ . \label{eq:SupertransFromFull}
\end{eqnarray}
This is exactly the expression found in the Pauli-Fierz theory both for translations (equation (\ref{eq:TransFromFull})) 
and pure supertranslations (equation (\ref{eq:SupertransFromFull})).

We have explicitly subtracted the zero mode of $T$ and the first spherical harmonic of $W$ (if any) to explicitly enforce the quotient $T \sim T + c$ (where $c$ is a constant) and $W \sim W + u_1$ (where $u_1$ is a linear combination of the $Y_{1}^m$), but we could have kept these terms in equation (\ref{eq:SupertransFromFull}) since the quotient is in fact automatically implemented in the formula for the generators of the pure supertranslations due to the above identities.

We thus see that the supertranslations of the full Einstein theory split into their zero modes (ordinary spacetime translations) and the rest (pure supertranslations) as one  linearizes the theory.  From the point of view of the linearized theory, these have a different origin. Spacetime translations are rigid global symmetries of standard type and their canonical generators have a pure bulk expression.  By contrast, pure supertranslations are improper gauge symmetries of the free spin-$2$ theory, and their canonical generators are thus given by a bulk expression proportional to the constraints, supplemented by a non-vanishing surface integral at infinity.   On-shell, these generators reduce to surface integrals.

We now turn to the angular momentum, given in the full Einstein theory by \cite{Henneaux:2018hdj,Henneaux:2019yax}
\be
\frac12 b_{ij} M^{ij} = \frac12 b_{ij} \int d^3 x x^{[j} \mathcal{G}^{i]}_{\textrm{Full}} + \frac12 b^{ij} \int_{S_2^{\infty}} d\theta d\varphi  2Y^{A}_{ij} (\bar{\pi}^{rB}_{\textrm{Full}}\bar{h}^{\textrm{Full}}_{AB} +  \xbar \gamma_{AB} \pi_{\textrm{Full}}^{(2)rB})
\ee
where the superscript $(2)$ refers to the asymptotic expansion, and denotes the coefficient of the next-to-leading term in  ${\pi}^{rB}_{\textrm{Full}}$ (coefficient of $\frac{1}{r^2}$ term).  The surface integral containing $\pi_{\textrm{Full}}^{(2)rB}$, which is linear in the field, can be conveniently rewritten in cartesian coordinates as $ \frac12 b_{ij} \oint_{S_2^{\infty}} d^2S_l 2 x^{[j}\pi_{\textrm{Full}}^{(2)i]l}$, or
\be
 \frac12 b_{ij} \oint_{S_2^{\infty}} d^2S_l 2 x^{[j}\pi_{\textrm{Full}}^{i]l},
 \ee
which involves  also the leading term in the asymptotic expansion and not just $\pi_{\textrm{Full}}^{(2)rB}$.  The would-be  divergence associated with this leading term is actually absent thanks to the boundary conditions \cite{Henneaux:2018hdj,Henneaux:2019yax}.
 
Expanding the angular momentum generator in powers of $\kappa$, one finds again that the term of order one,
$$
\frac12 b_{ij}\Bigg[ \int d^3 x x^{[j} \mathcal{G}^{i]}[\pi^{mn}] + 2 \oint_{S_2^{\infty}} d^2S_l  x^{[j}\pi^{i]l} \Bigg] \ ,
$$ identically vanishes due to the identity
$$
2\oint_{S_2^{\infty}} d^2S_l  x^{[j}\pi^{i]l} = 2\int d^3 x \partial_l(x^{[j}\pi^{i]l})= 2\int d^3 x (\pi^{[ij]} +  x^{[j} \partial_l \pi^{i]l})= -\int d^3 x x^{[j} \mathcal{G}^{i]}[\pi^{mn}]
$$
while the term of order two,
$$
\frac12 b_{ij}\Bigg[ \int d^3 x x^{[j} (\mathcal{G}^{i]}[p^{mn}] + \mathcal{P}^{i]}) + 2 \oint_{S_2^{\infty}} d^2S_l  x^{[j}p^{i]l}   +  \int_{S_2^{\infty}} d\theta d\varphi  2Y^{Aij} (\bar{\pi}^{rB}\bar{h}_{AB} )\Bigg]
$$
can be transformed into
\be
\frac12 b_{ij}\Bigg[ \int d^3 x x^{[j}  \mathcal{P}^{i]}  +  \int_{S_2^{\infty}} d\theta d\varphi  2Y^{Aij} (\bar{\pi}^{rB}\bar{h}_{AB} ) \Bigg]
\ee
using the same identity but with $\pi^{ij} \rightarrow p^{ij}$.  This is again exactly the expression of the angular momentum of the Pauli-Fierz theory.

Finally, the boost charges can be treated similarly.  Agreement of the weak field expansion of the boost charges with the expression derived within the linear theory is also found in that case, although the derivation is more cumbersome.

\section{Conclusions}
\label{sec:Conclusions}

In this paper, we have worked out the asymptotic structure of the linearized Einstein theory.  We have shown that it is described by the same infinite-dimensional symmetry group as the full interacting theory.  In that respect,   the gravitational field behaves at spatial infinity as its free limit, contrary what has been uncovered in \cite{Tanzi:2020fmt} for the Yang-Mills field (see also \cite{Christodoulou1981} in that context).

Some of the symmetries, which are all improper gauge symmetries in the full Einstein theory, appear in the linear theory as rigid symmetries with a non-trivial bulk contribution to their charge-generators (Poincar\'e transformations), while the others  appear as improper gauge symmetries with a generator that reduces on-shell to a surface integral at infinity, with no bulk contribution (pure supertranslations).   

Central in the analysis is the role played by the boosts.  The requirement that the boosts act as bona fide symmetries -- in particular, are canonical transformations -- naturally leads to the condition $\xbar h_{rA} = 0$ on the leading order of the mixed radial-angular components of the Pauli-Fierz field, as in the Einstein theory. It is that condition that dictates the size of the set of pure supertranslations, which depend on a single function of the angles, and not on four. The role played by the boosts raises the interesting question as to whether non-relativistic theories of gravity might have a bigger group of asymptotic symmetries.

The same study can be pursued for anti-de Sitter gravity, by linearizing the gravitational action around the anti-de Sitter background.  With the boundary conditions inherited from \cite{Henneaux:1985tv}, one finds that there is no non-trivial improper gauge symmetries (i.e., asymptotic symmetries with non-vanishing charge-generator), and that the only symmetries are the background $SO(3,2)$ symmetries.  This is in line with the group theoretical obstructions described in \cite{Safari:2019zmc}.   This negative result holds for standard symmetries of the theory, which leave the action invariant and which are then described by Lie algebras.  Going beyond that context would open new possibilities \cite{Compere:2019bua,Compere:2020lrt}.

The fact that the free Pauli-Fierz theory and the Einstein theory share similar asymptotic behaviours at spatial infinity can be used to investigate in the simpler Pauli-Fierz case various asymptotic questions such as (i) the 
incorporation of $\log r$ supertranslations; and  (ii) the role of super-rotations \cite{Banks:2003vp,Barnich:2010eb,Barnich:2009se}.  Our analysis could also be 
generalized to higher spin gauge fields \cite{Campoleoni:2017mbt}.  Work on these issues is currently in progress. 

\section*{Acknowledgements}
O.~F. holds a ``Marina Solvay'' fellowship. This work was partially supported by the ERC Advanced Grant ``High-Spin-Grav'',  by FNRS-Belgium (convention IISN 4.4503.15), as well as by funds from the Solvay Family.

\appendix

\section{Weak field expansion}
\label{App:Linear}

The weak-field expansion takes the form
\be
g_{ij} = \delta_{ij} + \kappa h^{ij} + \kappa^2  \varphi^{ij} +O(\kappa^3), \quad \pi^{ij}_{\textrm{Full}} = \kappa \pi^{ij} + O(\kappa^2) , \quad N = 1 + \kappa n + O(\kappa^2), \quad  N^i =  \kappa n^i + O(\kappa^2) \label{eq:WeakF}
\ee
where $\kappa$ is a small parameter.  The linear theory is obtained by keeping  the terms of order $\kappa^2$.   

The expansion goes explicitly as follows: the action of \cite{Dirac:1958jc,Arnowitt:1962hi} reads $ \int dt \{ \int d^3x (\pi^{ij}_{\textrm{Full}} \dot{g}_{ij} - N \mathcal{H} - N^k \mathcal{H}_k) - B_\infty\}$ where $B_\infty$ is a surface term at spatial infinity, the ``ADM energy'', $B_\infty = \int d^2 S_k \mathcal{E}^k_{\textrm{ADM}}[g_{ij} - \delta_{ij}]$. The vector  $\mathcal{E}^k_{\textrm{ADM}}$ is linear in its argument and identically fulfills $\partial_k \mathcal{E}^k_{\textrm{ADM}}[\Phi_{ij}] = - \nabla^{i}\nabla^{j}\Phi_{ij}+\triangle \Phi$ for any symmetric tensor $\Phi_{ij}$  \cite{Arnowitt:1962hi}.  If one expands the Dirac-ADM action in powers of $\kappa$, one gets a linear term, $\kappa \int dt \{ \int d^3 x (- \nabla^{i}\nabla^{j}h_{ij}+\triangle h) - \int d^2 S_k \mathcal{E}^k_{\textrm{ADM}}[h_{ij}])\}$, which is identically zero according to the property of $\mathcal{E}^k_{\textrm{ADM}}$.  A similar cancellation occurs at order $\kappa^2$ between the bulk term $\int dt \{ \int d^3 x (- \nabla^{i}\nabla^{j}\varphi_{ij}+\triangle \varphi)$ and the surface term $-\int d^2 S_k \mathcal{E}^k_{\textrm{ADM}}[\varphi_{ij}])\}$ (which are now generically non zero on-shell\footnote{In fact, they are equal to (minus) the energy of the linearized theory, as it follows from the Hamiltonian constraint at the next order, $\mathcal{E} +  (- \nabla^{i}\nabla^{j}\varphi_{ij}+\triangle \varphi) = 0$.}), leaving as coefficient of the $\kappa^2$-term the action (\ref{eq:Action}). 

The Einstein-Hilbert action is invariant under diffeomorphisms (which preserve the asymptotic behaviour). 
We discuss here the form of these symmetries in the covariant language. 

The spacetime metric $g_{\mu \nu}$ transforms as follows under diffeomorphisms,
\be
\delta_\zeta g_{\mu \nu} = \mathcal{L}_\zeta g_{\mu \nu} = \zeta^\rho \partial_\rho g_{\mu \nu} + \partial_\mu \zeta^\rho g_{\rho \nu} + \partial_\nu \zeta^\rho g_{\mu \rho } \, .
\ee
Now, a general diffeomorphism can formally be expanded in powers of $\kappa$.  In order to be compatible with the weak field expansion of the metric, $g_{\mu \nu} = g_{\mu \nu}^{\textrm{Minkowski}} + \kappa h_{\mu \nu} + O(\kappa^2)$, the variation of the leading background term, which is non-dynamical, should be zero.  Indeed, it cannot be compensated for by a variation of the dynamical fields.   Thus the allowed diffeomorphisms take the form
\be
\zeta^\mu = \xi^\mu + \kappa \epsilon^\mu + O(\kappa^2)
\ee
where $ \xi^\mu$ is a Killing vector of the Minkowskian metric and thus an infinitesimal Poincar\'e transformation, and where $\epsilon^\mu$ is only restricted by the asymptotic conditions analysed in the text.

Under a diffeomorphism, the first-order term $h_{\mu \nu}$ in the $\kappa$-expansion transforms as
\be
\delta_\zeta h_{\mu \nu} = \mathcal{L}_\xi h_{\mu \nu}  + \mathcal{L}_\epsilon g_{\mu \nu}^{\textrm{Minkowski}} 
\ee
There are thus two types of symmetries, rigid Poincar\'e symmetries inherited from the background
\be
\delta_\xi h_{\mu \nu} = \mathcal{L}_\xi h_{\mu \nu} 
\ee
and gauge transformations of the form 
\be
\delta_\epsilon h_{\mu \nu} =  \mathcal{L}_\epsilon g_{\mu \nu}^{\textrm{Minkowski}}  = \nabla_\mu \epsilon_\nu + \nabla_\nu \epsilon_\mu \, . \label{eq:GaugeA0}
\ee
This is the starting point of the discussion in the text.

\section{A technical lemma}
\label{App0}

In this appendix, we recall well known and useful technical facts.

\vspace{.2cm}
\noindent
{\bf Lemma:}
Let $f$ be a function such that 
\be
\partial_i f = \frac{\mu_i (x^A)}{r} \label{eq:app0}
\ee
where $\partial_i $ is the derivative with respect to the cartesian coordinates $x^i$ and where $x^A$ denotes the angles.  Then 
\be
f = a \ln r + g(x^A)
\ee
where $a$ is a constant.  

\vspace{.2cm}
\noindent
{\bf Proof:} From (\ref{eq:app0}), one gets 
\be
\partial_r f = \frac{k(x^A)}{r}, \qquad \partial_B f = m_B(x^A)
\ee
where $k(x^A) \equiv n^i \mu_i$ and $$m_B(x^A) \equiv \frac{\partial x^i}{\partial x^B} \frac{\mu_i (x^A)}{r}$$
are functions that depend only on the angles and not on $r$. The integrability condition that second derivatives should commute imposes $\partial_A k = 0$, i.e., $k = a$ with $a$ a constant.  A direct integration leads then to the desired result, where the integration constant can be absorbed in a redefinition of $g(x^A)$.

\vspace{.2cm}

It should be noted that things are simpler if $\frac{1}{r}$ is replaced by $\frac{1}{r^2}$ in (\ref{eq:app0}).  A direct reasoning shows indeed that then $f = a + O(\frac{1}{r})$ where $a $ is a constant.  Thus, if (\ref{eq:app0}) is replaced by $ \partial_i f = \frac{\mu_i (x^A)}{r} + O(\frac{1}{r^2})$, one finds $f = a \ln r + g(x^A) + O(\frac{1}{r})$.

\section{Transformation laws of the subleading orders  under Poincar\'e transformations\label{AppA}}

The transformation laws under
Poincar\'e transformations of the subleading terms in the asymptotic expansion of the metric and its conjugate momentum  are given by
\begin{eqnarray}
\delta h_{rr}^{(2)} & = & \frac{b}{\sqrt{\bar{\gamma}}}\left(\pi_{(2)}^{rr}-\pi_{(2)A}^{A}\right)+\frac{a^{0}}{\sqrt{\bar{\gamma}}}\left(\bar{\pi}^{rr}-\bar{\pi}_{A}^{A}\right)+Y^{A}\partial_{A}h_{rr}^{(2)}-w_{1}\bar{h}_{rr}+\partial_{A}w_{1}\partial^{A}\bar{h}_{rr}\,,\\
\nonumber \\
\delta h_{rA}^{(2)} & = & \mathcal{L}_{Y}h_{rA}^{(2)}+\frac{2b}{\sqrt{\bar{\gamma}}}\pi_{(2)A}^{r}+\frac{2a^{0}}{\sqrt{\bar{\gamma}}}\bar{\pi}_{A}^{r}+\partial_{A}w_{1}\bar{h}_{rr}-\partial_{B}w_{1}\bar{h}_{A}^{B}\,,\\
\nonumber \\
\delta h_{AB}^{(2)} & = & \mathcal{L}_{Y}h_{AB}^{(2)}+\frac{b}{\sqrt{\bar{\gamma}}}\left[2\pi_{AB}^{(2)}-\bar{\gamma}_{AB}\left(\pi_{(2)}^{rr}+\pi_{(2)C}^{C}\right)\right] \nonumber \\
 &  & +\frac{a^{0}}{\sqrt{\bar{\gamma}}}\left[2\bar{\pi}_{AB}-\bar{\gamma}_{AB}\left(\bar{\pi}^{rr}+\bar{\pi}_{C}^{C}\right)\right]+w_{1}\bar{h}_{AB}+\partial_{C}w_{1}\partial^{C}\bar{h}_{AB}+2\partial^{C}\partial_{(A}w_{1}\bar{h}_{B)C}\,,\\
\nonumber \\
\delta\pi_{(2)}^{rr} & = & \mathcal{L}_{Y}\pi_{(2)}^{rr}+\sqrt{\bar{\gamma}}\Big[\frac{1}{2}b\left(8h_{rr}^{(2)}+2h^{(2)}+4\bar{D}_{A}h_{(2)}^{rA}+\bar{\triangle}h_{rr}^{(2)}\right)\nonumber \\
 &  & +\bar{D}_{A}b\Big(2h_{(2)}^{rA}-\frac{1}{2}\bar{D}^{A}h^{(2)}+\bar{D}_{B}h_{(2)}^{AB}\Big)+\frac{1}{2}a^{0}\left(2\bar{h}_{rr}+\bar{\triangle}\bar{h}_{rr}\right)\Big]\nonumber \\
 &  & -2\partial_{A}w_{1}\bar{\pi}^{rA}+\partial^{A}w_{1}\partial_{A}\bar{\pi}^{rr}+\xbar \triangle w_{1}\bar{\pi}^{rr}\,,\\
\nonumber \\
\delta\pi_{(2)}^{rA} & = & \mathcal{L}_{Y}\pi_{(2)}^{rA}+\sqrt{\bar{\gamma}}\Big[\frac{1}{2}b\left(3h_{(2)}^{rA}+2\bar{D}_{B}h_{(2)}^{AB}+\bar{\triangle}h_{(2)}^{rA}-2\bar{D}^{A}h_{rr}^{(2)}-2\bar{D}^{A}h^{(2)}-\bar{D}^{A}\bar{D}_{B}h_{(2)}^{rB}\right) \nonumber \\
 &  & +\frac{1}{2}\bar{D}_{B}b\left(2h_{(2)}^{AB}+\bar{D}^{B}h_{(2)}^{rA}-\bar{D}^{A}h_{(2)}^{rB}\right)+\frac{1}{2}a^{0}\left(\bar{D}_{B}\bar{h}^{AB}-\bar{D}^{A}\bar{h}_{rr}-\bar{D}^{A}\bar{h}\right)\Big] \nonumber \\
 &  & -w_{1}\bar{\pi}^{rA}+\partial^{A}w_{1}\bar{\pi}^{rr}-\partial_{B}w_{1}\bar{\pi}^{AB}+\partial^{B}w_{1}\partial_{B}\bar{\pi}^{rA}-\partial^{A}\partial_{B}w_{1}\bar{\pi}^{rB}+\xbar \triangle w_{1}\bar{\pi}^{rA}\,,\\
\nonumber \\
\delta\pi_{(2)}^{AB} & = & \mathcal{L}_{Y}\pi_{(2)}^{AB}+\sqrt{\bar{\gamma}}\Big\{\frac{1}{2}b\left(\bar{\triangle}h_{(2)}^{AB}-2\bar{D}^{(A}\bar{D}_{C}h_{(2)}^{B)C}+\bar{D}^{A}\bar{D}^{B}h_{rr}^{(2)}+\bar{D}^{A}\bar{D}^{B}h^{(2)}\right) \nonumber \\
 &  & -\frac{1}{2}\bar{D}_{C}b\left[-\bar{D}^{C}h_{(2)}^{AB}+2\bar{D}^{(A}h_{(2)}^{B)C}+\gamma^{AB}\left(\bar{D}^{C}h_{rr}^{(2)}+\bar{D}^{C}h^{(2)}-2\bar{D}_{F}h_{(2)}^{CF}\right)\right] \nonumber \\
 &  & +\frac{1}{2}a^{0}\left[-2\bar{h}^{AB}+\bar{\gamma}^{AB}\left(\bar{h}-\bar{h}_{rr}\right)+\bar{\triangle}\bar{h}^{AB}-2\bar{D}^{(A}\bar{D}_{C}\bar{h}^{B)C}+\bar{D}^{A}\bar{D}^{B}\bar{h}_{rr}+\bar{D}^{A}\bar{D}^{B}\bar{h}\right]\Big\} \nonumber \\
 &  & -2w_{1}\bar{\pi}^{AB}+2\partial^{(A|}w_{1}\bar{\pi}^{r|B)}+\partial^{C}w_{1}\partial_{C}\bar{\pi}^{AB}-2\partial^{(A}\partial_{C}w_{1}\bar{\pi}^{B)C}+\xbar \triangle w_{1}\bar{\pi}^{AB}\,,
\end{eqnarray}
where
\begin{eqnarray}
\mathcal{L}_{Y}h_{rA}^{(2)} & = & Y^{B}\partial_{B}h_{rA}^{(2)}+\partial_{A}Y^{B}h_{rB}^{(2)}\, , \\
\mathcal{L}_{Y}h_{AB}^{(2)} & = & Y^{C}\partial_{C}h_{AB}^{(2)}+\partial_{A}Y^{C}h_{BC}^{(2)}+\partial_{B}Y^{C}h_{AC}^{(2)}\,,\\
\mathcal{L}_{Y}\pi_{(2)}^{rr} & = & \partial_{A}\left(Y^{A}\pi_{(2)}^{rr}\right)\,,\\
\mathcal{L}_{Y}\pi_{(2)}^{rA} & = & -\partial_{B}Y^{A}\pi_{(2)}^{rB}+\partial_{B}\left(Y^{B}\pi_{(2)}^{rA}\right)\,,\\
\mathcal{L}_{Y}\pi_{(2)}^{AB} & = & -\partial_{C}\xi^{A}\pi_{(2)}^{BC}-\partial_{C}\xi^{B}\pi_{(2)}^{AC}+\partial_{C}\left(\xi^{C}\pi_{(2)}^{AB}\right)\,.
\end{eqnarray}

\section{Absence of divergences in bulk piece of homogeneous Lorentz transformations}
\label{App:NoDiv}

\subsection{Absence of divergence in spatial rotations}

The Poincar\'e Killing vectors of spatial rotations
read
\be
\xi = 0\, , \qquad \xi^{r}  =0\, , \qquad
\xi^{A}  =Y^{A}\, ,
\ee
from which, one sees that the bulk part of (\ref{eq:Gcan})
possesses the following potential logarithmic divergence
\begin{equation}
\int d^{3}x\xi^{i}\mathcal{P}_{i}=\int drd\theta d\varphi\left[\frac{1}{r}Y^{A}\mathcal{P}_{A}^{(1)}+\mathcal{O}(r^{-2})\right]\,,\label{eq:BulkMom1}
\end{equation}
with
\begin{equation}
\mathcal{P}_{A}^{(1)}=\bar{\pi}^{rr}\partial_{A}\bar{h}_{rr}+\bar{\pi}^{BC}\bar{D}_{A}\bar{h}_{BC}-2\bar{D}_{C}\left(\bar{\pi}^{BC}\bar{h}_{BA}\right)\,.
\end{equation}

The integral on the 2-sphere in (\ref{eq:BulkMom1}) can be rewritten
as
\begin{equation}
\int d\theta d\varphi Y^{A}\mathcal{P}_{A}^{(1)}=\int d\theta d\varphi\Big[(\bar{\pi}^{rr}-\bar{\pi}_{A}^{A})Y^{A}\partial_{A}\bar{h}_{rr}+\bar{\pi}^{AB}\mathcal{L}_{Y}\bar{k}_{AB}+2Y_{A}\bar{D}_{B}(\bar{h}_{rr}\bar{\pi}^{AB})\Big]\,.\label{eq:AngInt}
\end{equation}
where 
\be
\bar{k}_{AB}=\frac{1}{2}\left(\bar{h}_{AB}+\bar{h}_{rr}\bar{\gamma}_{AB}\right).
\ee
The parity conditions on the asymptotic fields imply
that $(\bar{\pi}^{rr}-\bar{\pi}_{A}^{A})$ is strictly odd and that
the field $\bar{k}_{AB}$ is subject to the following parity condition
\begin{align}
\bar{k}_{AB} & =(\bar{k}_{AB})^{\text{even}}+\bar{D}_{A}\bar{D}_{B}U+\bar{\gamma}_{AB}U\,.\label{eq:PCkAB}
\end{align}

It is then straightforward to see that the first term in the integrand
of (\ref{eq:AngInt}) is odd and so vanishes upon integration on the 2-sphere. The third
term also vanishes  recalling that
$\bar{D}_{(A}Y_{B)}=0$ (one can freely integrate by parts on the boundaryless $2$-sphere). Thus, (\ref{eq:AngInt})  reduces to its second term, which becomes, by decomposing the asymptotic fields into their parity components,
\begin{align}
\int d\theta d\varphi Y^{A}\mathcal{P}_{A}^{(1)} & =\int d\theta d\varphi\Big[(\bar{\pi}^{AB})^{\text{odd}}\mathcal{L}_{Y}(\bar{k}_{AB})^{\text{even}}+\sqrt{\bar{\gamma}}(\bar{D}^{A}\bar{D}^{B}V-\bar{\gamma}^{AB}\bar{\triangle}V)\mathcal{L}_{Y}(\bar{k}_{AB})^{\text{even}} \nonumber \\
 & \quad+(\bar{\pi}^{AB})^{\text{odd}}\mathcal{L}_{Y}\Big(\bar{D}_{A}\bar{D}_{B}U+\bar{\gamma}_{AB}U\Big) \nonumber \\ & +\sqrt{\bar{\gamma}}(\bar{D}^{A}\bar{D}^{B}V-\bar{\gamma}^{AB}\bar{\triangle}V)\mathcal{L}_{Y}\Big(\bar{D}_{A}\bar{D}_{B}U+\bar{\gamma}_{AB}U\Big)\Big]\,.
\end{align}
The first and fourth terms in the above integral turn out to be both
odd, vanishing upon integration on the 2-sphere, while the remaining
terms (second and third) are both even. Integrating by parts, we get
that
\begin{align}
\int d\theta d\varphi Y^{A}\mathcal{P}_{A}^{(1)} & =\int d\theta d\varphi\Big[\sqrt{\bar{\gamma}}\mathcal{L}_{Y}\Big(\bar{D}^{A}\bar{D}^{B}(\bar{k}_{AB})^{\text{even}}-\bar{\triangle}(\bar{k}_{A}^{A})^{\text{even}}\Big)V \nonumber \\ & +\mathcal{L}_{Y}\Big(\bar{D}_{A}\bar{D}_{B}(\bar{\pi}^{AB})^{\text{odd}}+\bar{\gamma}_{AB}(\bar{\pi}^{AB})^{\text{odd}}\Big)U\Big]\,,\label{eq:AngInt2}
\end{align}
where we used the fact that the Lie derivative along the spatial
rotations $Y^{A}$ commutes with the covariant derivative on the 2-sphere. It
is then clear that (\ref{eq:AngInt2}) vanishes by virtue of the leading
orders of the constraint equations (\ref{eq:Const1}), (\ref{eq:Const2}),
and(\ref{eq:Const3}), which lead to the following equations
\begin{align}
\bar{D}_{A}\bar{D}_{B}\bar{\pi}^{AB}+\bar{\pi}_{A}^{A} & = 0\,,\label{eq:Const1-2}\\
\bar{D}^{A}\bar{D}^{B}\bar{k}_{AB}-\bar{\triangle}\bar{k}_{A}^{A} &  = 0\,.\label{eq:Const2-2}
\end{align}
This shows that the logarithmic divergence is in fact absent in the bulk integral for the angular momentum.

\subsection{Absence of divergence in the boost generator }

The boost  Killing vector reads
\begin{equation}
\xi=br\,, \qquad \xi^i = 0 \, ,
\end{equation}
which leads to the following potential logarithmic divergence  in
the leading bulk part of the canonical generator (\ref{eq:Gcan})
\begin{equation}
\int d^{3}x\,\xi\mathcal{E}=\int drd\theta d\varphi\left[\frac{1}{r}b\mathcal{E}^{(2)}+\mathcal{O}(r^{-2})\right]\,,\label{eq:BulkEn1}
\end{equation}
where
\begin{eqnarray}
\mathcal{E}^{(2)} & = & \frac{1}{\sqrt{\bar{\gamma}}}\Bigg(\frac{1}{2}\bar{\pi}_{rr}^{2}-\bar{\pi}_{rr}\bar{\pi}+2\bar{\pi}_{A}^{r}\bar{\pi}^{rA}+\bar{\pi}^{AB}\bar{\pi}_{AB}-\frac{1}{2}\bar{\pi}{}^{2}\Bigg)+\sqrt{\bar{\gamma}}\Bigg[-\bar{h}_{rr}^{2}+\frac{3}{2}\bar{h}_{rr}\bar{h}_{A}^{A} \nonumber \\
 &  & +\frac{3}{4}\bar{h}_{AB}\bar{h}^{AB}+\frac{3}{4}\bar{h}^{2}+\frac{1}{2}\bar{D}_{A}\bar{h}_{rr}\bar{D}^{A}\bar{h}_{rr}+\frac{1}{2}\bar{D}_{A}\bar{h}_{rr}\bar{D}^{A}\bar{h}+\bar{D}_{A}\Big(\bar{h}_{rr}\bar{D}_{B}\bar{h}^{AB}\Big) \nonumber \\
 &  & +\frac{1}{4}\bar{D}_{C}\bar{h}_{AB}\bar{D}^{C}\bar{h}^{AB}-\frac{1}{2}\bar{D}_{A}\bar{h}^{AB}\bar{D}_{C}\bar{h}_{B}^{\,\,\,C}+\frac{1}{4}\bar{D}_{A}\bar{h}\bar{D}^{A}\bar{h} \nonumber \\
 & & +\bar{D}_{A}\Big(\bar{h}\bar{D}_{B}\bar{h}^{AB}+\bar{h}^{AB}\bar{D}_{C}\bar{h}_{B}^{\,\,\,C}\Big)\Bigg]\,.
\end{eqnarray}
Turning to the variables $\xbar k_{AB}$, 
the integral on the 2-sphere in (\ref{eq:BulkEn1}) can be
rewritten as
\begin{align}
\int d\theta d\varphi\,b\mathcal{E}^{(2)} & =\int d\theta d\varphi\,b\Bigg[\frac{1}{\sqrt{\bar{\gamma}}}\left(\frac{1}{2}\bar{\pi}_{rr}^{2}-\bar{\pi}_{rr}\bar{\pi}+2\bar{\pi}_{A}^{r}\bar{\pi}^{rA}+\bar{\pi}^{AB}\bar{\pi}_{AB}-\frac{1}{2}\bar{\pi}{}^{2}\right)\label{eq:MomBulkEn}\\
 & \qquad\qquad\qquad+\sqrt{\bar{\gamma}}\Big(-2\bar{h}_{rr}^{2}-\frac{1}{2}\bar{h}_{rr}\bar{\triangle}\bar{h}_{rr}+3\bar{k}_{AB}\bar{k}^{AB}+3\bar{k}^{2} \nonumber \\
 & \qquad\qquad\qquad+4\bar{k}^{AB}\bar{D}_{A}\bar{D}_{C}\bar{k}_{B}^{C}+\bar{D}_{A}\bar{k}\bar{D}^{A}\bar{k} +2\bar{D}_{A}\bar{k}^{AB}\bar{D}_{C}\bar{k}_{B}^{C}\nonumber\\
 & \qquad\qquad\qquad\qquad\quad\quad+4\bar{D}^{A}\bar{k}\bar{D}_{B}\bar{k}_{A}^{B}+4\bar{k}\bar{D}_{A}\bar{D}_{B}\bar{k}^{AB}+\bar{D}_{C}\bar{k}_{AB}\bar{D}^{C}\bar{k}^{AB}\Big)\Bigg]\,.\label{eq:En2}
\end{align}
where $\bar{\pi}=\bar{\pi}_{A}^{A}$ and $\bar{k}=\bar{k}_{A}^{A}$.

First, let us focus on the terms containing quadratic contributions
in the momentum components (\ref{eq:MomBulkEn})

\begin{equation}
I_{\pi}=\int d\theta d\varphi\,\frac{b}{\sqrt{\bar{\gamma}}}\Big(\frac{1}{2}\bar{\pi}_{rr}^{2}-\bar{\pi}_{rr}\bar{\pi}+2\bar{\pi}_{A}^{r}\bar{\pi}^{rA}+\bar{\pi}^{AB}\bar{\pi}_{AB}-\frac{1}{2}\bar{\pi}{}^{2}\Big)\,.\label{eq:Ipi}
\end{equation}
By making use of the parity conditions,
we can neglect the odd contributions to (\ref{eq:Ipi}), which vanish
upon integration on the 2-sphere. Thus, up to some integration by
parts, (\ref{eq:Ipi}) can be reduced as follows
\begin{align}
I_{\pi} & =\int d\theta d\varphi\,b\Big[-4(\bar{\pi}^{rA})^{\text{even}}\bar{D}_{A}V+2(\bar{\pi}^{AB})^{\text{odd}}\bar{D}_{A}\bar{D}_{B}V\Big]\label{eq:Ipi2}\\
 & =\int d\theta d\varphi\,\Big\{4\bar{D}_{A}b\left[\bar{D}_{B}(\bar{\pi}^{AB})^{\text{odd}}+(\bar{\pi}^{rA})^{\text{even}}\right]\nonumber \\
 & \qquad\qquad\quad+b\left[4\bar{D}_{A}(\bar{\pi}^{rA})^{\text{even}}+2\bar{D}_{A}\bar{D}_{B}(\bar{\pi}^{AB})^{\text{odd}}\right]+2\bar{D}_{A}\bar{D}_{B}b(\bar{\pi}^{AB})^{\text{odd}}\Big\} V\,.\label{eq:Ipi3}
\end{align}
By considering the equation for the boost parameter
\begin{equation}
\bar{D}_{A}\bar{D}_{B}b+\bar{\gamma}_{AB}b=0\,,
\end{equation}
and the constraint equation (\ref{eq:Const1-2}), the integral (\ref{eq:Ipi3})
becomes
\begin{equation}
I_{\pi}=4\int d\theta d\varphi\,\Big\{\bar{D}_{A}b\left[\bar{D}_{B}(\bar{\pi}^{AB})^{\text{odd}}+(\bar{\pi}^{rA})^{\text{even}}\right]+b\left[\bar{D}_{A}(\bar{\pi}^{rA})^{\text{even}}-\bar{\pi}{}^{\text{odd}}\right]\Big\}\approx0\,,
\end{equation}
which vanishes by virtue of (\ref{eq:Const1}) and (\ref{eq:Const2}).

Now, we can focus in the remaining part  (\ref{eq:En2}), which is
quadratic in the fields $\bar{h}_{rr}$, $\bar{k}_{AB}$ and
their derivatives. We can immediately neglect the terms that are quadratic
in $\bar{h}_{rr}$, which turn out to be odd (multiplied by the odd
term $b$), and then vanish upon integration. Decomposing into parity components the field $\bar{k}_{AB}$ according to (\ref{eq:PCkAB}), using the
constraint equation (\ref{eq:Const2-2}), and integrating by parts,
we get that
\begin{align}
\int d\theta d\varphi\,b\mathcal{E}^{(2)} & =2\int d\theta d\varphi\sqrt{\bar{\gamma}}\Bigg\{\bar{D}_{A}b\Big(\bar{D}^{A}\bar{k}^{\text{even}}-[\bar{\triangle},\bar{D}^{A}]\bar{k}^{\text{even}}\Big) \nonumber \\
& \qquad \qquad \qquad +b\left(\bar{D}^{A}[\bar{\triangle},\bar{D}^{B}](\bar{k}_{AB})^{\text{even}}+[\bar{D}^{A},\bar{\triangle}]\bar{D}^{B}(\bar{k}_{AB})^{\text{even}}\right)\Bigg\} U\,.\label{eq:En3}
\end{align}
The first term in (\ref{eq:En3}) vanishes by virtue of the identity
$[\bar{\triangle},\bar{D}^{A}]\bar{k}^{\text{even}}=\bar{D}^{A}\bar{k}^{\text{even}}$
on the unit 2-sphere. For the second term we make use of the commutators
\begin{align}
[\bar{\triangle},\bar{D}^{B}](\bar{k}_{AB})^{\text{even}} & =2\bar{D}^{A}\bar{k}^{\text{even}}-3\bar{D}_{B}(\bar{k}^{AB})^{\text{even}}\,,\\{}
[\bar{D}^{A},\bar{\triangle}]\bar{D}^{B}(\bar{k}_{AB})^{\text{even}} & =\bar{D}^{A}\bar{D}^{B}(\bar{k}_{AB})^{\text{even}}\,.
\end{align}
We then obtain that (\ref{eq:En3}) reduces to
\begin{equation}
\int d\theta d\varphi\,b\mathcal{E}^{(2)}=-2\int d\theta d\varphi b\left[\bar{D}^{A}\bar{D}^{B}(\bar{k}_{AB})^{\text{even}}-\bar{\triangle}\bar{k}^{\text{even}}\right]\,,
\end{equation}
which is clearly zero since the condition (\ref{eq:Const2-2})
holds.
Thus, the potential logarithmic divergence is actually also absent in the bulk integral for the boost generator.

\end{document}